\documentclass[12pt]{iopart}

\usepackage{color}
\usepackage{psfrag}
\usepackage[pdftex]{graphicx}
\usepackage{epsfig}
\begin{document}

\textbf{[August 23, 2012]}

\title{Statistics of anomalously localized states at the center of band $E=0$ in the
one-dimensional Anderson localization model}

\author{V.E.Kravtsov$^{1,2}$ and V.I.Yudson$^{3}$}

\address{$^1$The Abdus Salam International Centre for
Theoretical Physics, P.O.B. 586, 34100 Trieste, Italy.\\
$^{2}$Landau Institute for Theoretical Physics, 2 Kosygina
st., 117940 Moscow, Russia.\\$^{3}$Institute for Spectroscopy,
Russian Academy of Sciences, 142190 Troitsk, Moscow, Russia}

\begin{abstract} 
We consider the distribution function $P(|\psi|^{2})$ of the
eigenfunction amplitude at the center-of-band ($E=0$) anomaly
in the one-dimensional tight-binding chain with weak uncorrelated
on-site disorder (the one-dimensional Anderson model). The special emphasis
is on the probability of the anomalously localized states (ALS) with
$|\psi|^{2}$ much larger than the inverse typical localization
length $\ell_{0}$. Using the solution to the generating function
$\Phi_{an}(u,\phi)$ found recently in our works \cite{KY-Ann,
KY-PRB} we find the ALS probability distribution $P(|\psi|^{2})$ at
$|\psi|^{2}\ell_{0}\gg 1$. As an auxiliary preliminary step we
found the asymptotic form of the generating function
$\Phi_{an}(u,\phi)$ at $u\gg 1$ which can be used to compute other
statistical properties at the center-of-band anomaly. We show that
at moderately large values of $|\psi|^{2}\ell_{0}$, the probability
of ALS at $E=0$ is
smaller than at energies away from the anomaly. However, at very
large values of $|\psi|^{2}\ell_{0}$, the tendency is inverted:
it is exponentially
easier to create a very strongly localized state at $E=0$ than at
energies away from the anomaly. We also found the leading term in the
behavior of $P(|\psi|^{2})$ at small $|\psi|^{2}\ll \ell_{0}^{-1}$
and show that it is consistent with the exponential localization
corresponding to the Lyapunov exponent found earlier by Kappus and
Wegner \cite{Wegner} and Derrida and Gardner \cite{Derrida}.

\end{abstract}

%
%
%
%
%
%
%
%

\pacs{72.15.Rn, 72.70.+m, 72.20.Ht, 73.23.-b}

\section{Introduction}\label{Intro}
There is a long-lasting interest in localization effects
\cite{Anderson,Abrahams} in 1d systems \cite{Borland}-\cite{KY-Ann}.
The simplest and most widely studied model is a linear chain with
a nearest-neighbor hopping
and random site energies $\varepsilon_i$ with no inter-site
correlation: $\langle \varepsilon_i\varepsilon_j\rangle =
\delta_{ij}\sigma^2$. The wave function $\psi_i$ at a site $i$
of this one-dimensional Anderson localization model \cite{Anderson}
obeys the equation:
\begin{equation}\label{1dAnderson}
\psi_{i-1} +\psi_{i+1} + \varepsilon_i\psi_{i} = E\psi_{i}.
\end{equation}
In the absence of disorder ($\varepsilon_i \equiv 0$) the
eigenstates would be plane waves, with eigenenergies determined by
the wave vector $k$: $E(k) = 2\cos(k)$, $k \in (-\pi, \pi)$. In the
presence of the disorder, the  eigenstates are random and require 
statistical description. Moreover, the states are {\it localized}
at an arbitrary small disorder strength
$\sigma$. For weak disorder the localization length $\ell(E)$ is
large as compared to the lattice constant: $\ell(E) \gg 1$. This
means that the ``typical" magnitude of the normalized wave function near
its localization center can be estimated as $|\psi|^2_{typ} \sim
1/\ell(E) \ll 1$. However, for some realizations of the disorder,
more strongly localized states, "anomalously localized states"
(ALS), are possible, with the value of the wave function maximum in
the range of $1/\ell(E) \ll |\psi|^2 \leq 1$ (the right equality
would correspond to a state localized at a single lattice site). Our
aim in the present paper is to study the probability distribution
$P(|\psi|^2)$ of such strongly localized states in a long
\emph{weakly disordered} chain.

We will be especially interested in the statistics of ALS  in the
vicinity of the so-called Kappus-Wegner center-of-band ($E=0$,
$k=\pi/2$ ) anomaly \cite{Wegner}. This anomaly is a feature of a
discrete chain (it is absent in the continuum model) and originates
from the commensurability of the de Broglie wavelength and the
lattice constant. The anomaly manifests itself \cite{Wegner,Derrida}
in a sharp, \emph{finite} in the limit $\sigma\rightarrow 0$,
enhancement of the density of states (DoS) $\nu(E=0)$ and the
localization length $\ell(E=0)$ inside a very narrow energy window
(of the width $\sim \sigma^2$) around the band center $E=0$ as
compared to their values
\begin{equation}\label{nu0-and-l0}
\nu_0(E=0) \approx \frac{1}{2\pi} \,\,\, ; \,\,\,\,\,\,\,\,
\ell_0 \equiv \ell_0(E=0) = \frac{8}{\sigma^2}
 \,
\end{equation}
beyond this interval \cite{box-distribution}. In
particular, it was shown \cite{Derrida} that in the limit $\sigma\ll
1$:
\begin{equation}\label{l-to-l0}
\hspace{-2cm}
\frac{\nu(E=0)}{\nu_0(E\rightarrow 0)} =
\frac{4\sqrt{2}\pi^{3}}{\Gamma^{4}(1/4)} = 1.01508... \,\,\,\, ; \,\,\,
\frac{\ell^{{\rm ext}}(E=0)}{\ell_0(E=0)} =
\frac{1}{16\pi^2}\Gamma^4\left(\frac{1}{4}\right) = 1.0942...
\, .
\end{equation}
Here we have introduced the superscript ``$ext$" to emphasize that
the corresponding localization length $\ell^{{\rm
ext}}=1/[\Re \,\gamma(E)]$ is defined by the Lyapunov exponent $\gamma(E)$
and therefore characterizes the exponentially decaying {\it tails}
of localized wave functions; for this reason it will be referred to
as the ''\emph{extrinsic}'' localization length. Similar anomalies
have been found later \cite{Titov,AL} for other physical quantities
(like transmission and conductance), also related with the Lyapunov
exponent.

In contrast to this set of problems, the eigenfunction statistics
$P(|\psi|^2)$ may provide information about an ``intrinsic" spatial
structure of localized wave functions including the vicinity of the
center of localization. In particular, it allows to calculate the
"intrinsic" localization length $\ell^{{\rm int}}(E)=1/I_{2}(E)$,
where $I_{2}(E)=\sum_{i}|\psi_{i}(E)|^{4}$ is the inverse
participation ratio.

However, studying the statistical properties of \emph{normalized}
eigenfunctions is a considerably more difficult theoretical problem
than studying the Lyapunov exponent (the latter is related to
propagation of an external wave in a semi-infinite chain and is not
directly related with {\it eigenfunctions}).

The formalism for studying the eigenfunction statistics in a
disordered chain (see review \cite{Mirlin2000}), adapted recently
\cite{KY-Ann} to the case of the center-of-band anomaly, expresses
moments of the eigenfunction distribution in terms of a ``generating
function" $\Phi(u, \phi; E)$ of the two auxiliary variables. These
variables can be loosely interpreted \cite{Wegner,KY-Ann} as the
squared amplitude $u \sim |a_{j}|^2\ell_0$ and the ``phase" $\phi$
defined by a representation of eigenfunctions in the form: $\psi_{j}
= a_{j}\cos{ (kj + \phi_{j})}$ with slowly varying $a_{j}
> 0$ and $\phi_{j} \in (0, \pi)$.  The generating function $\Phi(u, \phi; E)$
allows one to calculate all local statistics of eigenfunctions. In
particular, it determines the inverse participation ratio (IPR)
$I_{2}$ and higher moments $I_{m}
=\sum_{j}\langle|\psi_{j}|^{2m}\rangle$, as well as the full
distribution function $P(|\psi|^2)$.
Also, the generating function $\Phi(u,\phi;E)$ determines (through a
\emph{nonlinear} integral relation Eq.(\ref{P-u-phi})) the joint
probability distribution $P(u,\phi;E)$ of the amplitude and the phase.
However, the relationship between the generating function
$\Phi(u,\phi;E)$ and the normalized distribution function ${\cal
P}(\phi;E) = \int du P(u,\phi;E)$ of the phase $\phi$
turns out to be remarkably simple \cite{KY-Ann}, it is given by the
limit $u \rightarrow 0$ of the generating function $\Phi(u, \phi;
E)$:
\begin{equation} \label{u=0}
\Phi(u=0, \phi; E) = \mathcal{P}(\phi;E) = 2P_{refl}(\theta; E)|_{\theta = 2\phi}
\, ,
\end{equation}
There is also a simple relationship between ${\cal P}(\phi;E)$ and
the probability distribution $P_{refl}(\theta; E)$ of the reflection
phase $\theta$ for a wave incident on a semi-infinite disordered
chain. It is given by the second equality in Eq.(\ref{u=0}). At weak
disorder the phase distribution $\mathcal{P}(\phi; E)$ is uniform in
the continuum model and outside the center-of-band anomaly but it
becomes a non-trivial function of $\phi$ at $E=0$ \cite{Derrida}.

A relative simplicity of calculation of such quantities as the
Lyaupunov exponent (and the extrinsic localization length
$\ell^{{\rm ext}}(E)$) and the DoS, $\nu(E)$, is due to the fact
that they can be expressed entirely in terms of the the probability
distribution $\mathcal{P}(\phi; E)$, i.e. involve the generating
function $\Phi(u=0, \phi; E)$ at $u=0$.   For instance, the DoS,
$\nu(E)$ is given by \cite{Wegner,KY-Ann}:
\begin{equation}\label{corr}
\frac{\nu(E)}{\nu_{0}(E)}=4\pi\int_{0}^{\pi/2}d\phi\,\cos^{2}(\phi)\,\,[\mathcal{P}(\phi; E)]^{2}.
\end{equation}
On the contrary, the complexity of the problem of local
eigenfunction statistics arises because it requires the full
generating function $\Phi(u, \phi; E)$  of the two variables $u$ and
$\phi$. In particular, the statistics of relatively rare anomalously
localized eigenstates of large peak amplitude $|\psi|^{2}\ell_{0}\gg
1$ which we will study in the present paper is determined by
$\Phi(u, \phi; E)$ at large values of the variable $u\gg 1$.
\begin{figure}[t]
\centering
\includegraphics[height=8cm,width=6.5cm]{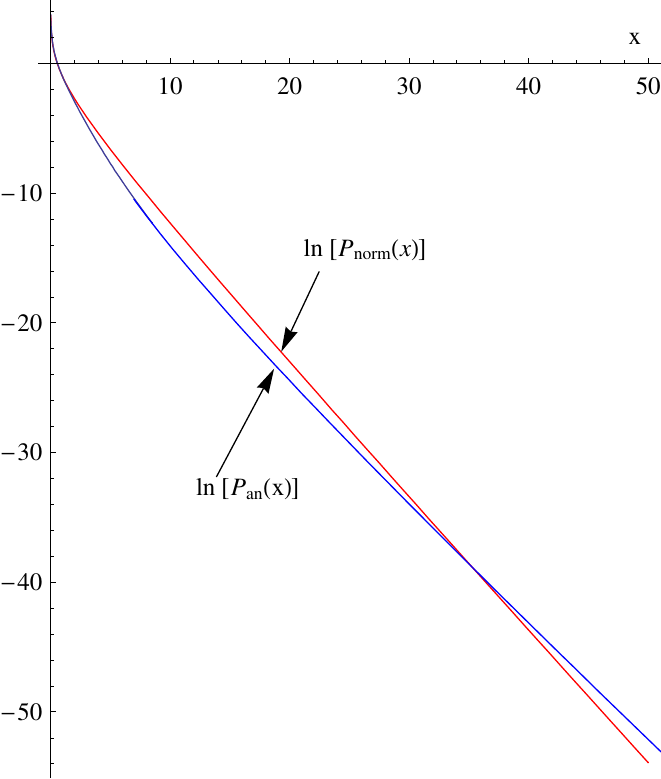}
\caption{The logarithm of the ALS probability distribution $P(x)$
($x=|\psi|^{2}\ell_{0})$ at the center-of-band anomaly $E=0$ and
outside ($E\neq 0$).   At moderately large values of
$|\psi|^{2}\ell_{0}$ ($|\psi|^{2}\ell_{0}<35$) the probability
of ALS at $E=0$ is smaller than
that for $E\neq 0$. However, at $|\psi|^{2}\ell_{0}>35$ the
situation is inverted: the probability ($\sim 10^{-15}$) of very
strongly localized states is larger at the band center. Note that at
weak disorder $\sigma \ll 1$ we consider in this paper the typical
localization length $\ell_{0}\approx 8/\sigma^{2}\gg 1$ is
parametrically larger than the lattice constant $a=1$ \cite{box-distribution}.
Thus the
above results are valid when the anomalously small localization
length is still much larger than the lattice constant.}
\label{tail-compar}
\end{figure}

The generating function $\Phi(u,\phi;E)$ for a long chain at the
center-of-band anomaly has been found recently \cite{KY-PRB, KY-Ann}
by solving exactly the corresponding second order partial differential
equation Eq.(\ref{anomal-equation}) in $u$ and $\phi$ variables. The exact
solution Eq.(\ref{canon-final}) to this equation  reflects a hidden
symmetry of the problem which has not been yet explicitly
exploited. However, the solution is given in quadratures as an
integral of a product of Whittaker functions
over the variable which enters both the argument and the index of these
functions. In this paper we perform a
careful analysis of the integral and derive the asymptotic form of
$\Phi(u,\phi)$ at large values of $u\gg 1$ (from now on we omit the
energy argument $E=0$ for brevity). It has a form:
\begin{equation}\label{Phi-an-asymp}
\Phi_{an}(u, \phi) = A(\phi) \frac{\mathrm{e}^{-\sqrt{u} \,
b(\phi)}}{u^{1/4}} \, \,\,\,\,; \,\,\, u \gg 1 \, ,
\end{equation}
where the function $b(\phi)$ is a solution to the first order
ordinary differential equation (\ref{Diff-Eq-for-b}), and $A(\phi)$
is specified in the section \ref{Asymptotic of Phi}.

It allows us to compute the tail of the distribution function
$P(|\psi|^{2})$ at $|\psi|^{2}\ell_{0}\gg 1$ in a long chain of the
length $L\gg \ell_{0}$:
\begin{equation}\label{P-an-asymp}
P_{an}(|\psi|^{2}) \sim  \frac{1} {\ell_{0}L}\,\frac{
\exp{(-\kappa\,|\psi|^2\ell_{0})}} {|\psi|^{6}} \, ,
\,\,\,\,\,\,\, ( |\psi|^{2}\ell_{0} \gg 1) \, ,
\end{equation}
where the coefficient $\kappa$ is determined by some ``critical
angle" $\phi_{c}$, given by
Eqs.(\ref{crit-angle-parametrization}),(\ref{crit-angle-nice}), at
which the function $b^{2}(\phi)/4\cos^{2}\phi$ reaches its minimum:
\begin{equation}\label{kappa}
\kappa= \frac{b^{2}(\phi_{c})}{4\cos^{2}\phi_{c}} = 0.830902... <
1.
\end{equation}

The anomalous distribution of eigenfunction amplitudes
Eq.(\ref{P-an-asymp}) should be compared with the ``normal" one
\cite{KY-Ann} valid in the continuum model and outside the
center-of-band anomaly in the discrete chain \cite{rem}:
\begin{equation}\label{1d-dist}
P_{norm}(|\psi|^{2}) =\frac{ \ell_{0}}{L}\,\frac{{\rm
exp}\left(-|\psi|^{2}\ell_{0}\right)}{|\psi|^{2}},
\,\,\,\,\,\,\,(|\psi|^{2}\ell_{0} \gg e^{-L/\ell_{0}}) \, .
\end{equation}
A comparison of Eqs.(\ref{P-an-asymp}) and (\ref{1d-dist}) reveals 
an unexpected feature (see Fig.1). While the probability of moderately
strongly localized states (with the peak intensity
$1<|\psi|^{2}\ell_{0}<35$) is smaller at $E=0$ than that away from
the anomaly, very strongly localized states (with
$|\psi|^{2}\ell_{0}>35$) are more probable at the band center.
Formally this re-entrant behavior is caused by the value of
$\kappa\approx 0.83 <1$ (see Eq.(\ref{kappa}))  at the "critical
angle" $\phi_{c}\neq 0$; for the "normal" case $E\neq 0$ one obtains
$b(\phi)=2$ and thus $\phi_{c}=0$ and $\kappa=1$.

The behavior of moderately strongly localized states is consistent
with the result Eq.(\ref{l-to-l0}) for the Lyapunov exponent which
gives an enhanced typical extrinsic localization length at $E=0$.
The opposite trend  for very strongly localized states is perhaps
due to the Bragg-mirror effect of the harmonics of the random
potential which double the period of the lattice \cite{Muz-Khm-95,Smol-Altsh-97}.

A point of special interest is the distribution of small amplitudes
$P(|\psi|^{2})$ at $|\psi|^{2}\ell_{0}\ll 1$, as it gives an idea on
the shape of the tail of the localized wave function. We found the
leading term $|\psi|^{-2}$ in $P(|\psi|^{2})$ at small
$|\psi|^{2}\ell_{0}$ and shown that it is universal for all systems
with exponentially localized eigenstates.

The rest of the paper is devoted to the derivation of the announced
results and is organized in the following way. In section
\ref{Asymptotic of Phi} we obtain the asymptotic of the generating
function $\Phi_{an}(u,\phi)$ at $u \gg 1$. In the subsequent section
\ref{Asymptotic of P} we derive the asymptotic of the probability
distribution function $P_{an}(|\psi|^2)$ at $|\psi|^2\ell_{0} \gg
1$. The behavior of $P(|\psi|)$ at small $|\psi|^{2}\ell_{0}$ is
analyzed in Sec.4. In the last section \ref{Conclusion} we
summarize and discuss the obtained results.

\section{Generating function $\Phi_{an}(u,\phi)$ and its asymptotic at $u \gg 1$}
\label{Asymptotic of Phi}
Sufficiently far from the ends of a long chain, the generating function
becomes site independent. At the center-of-band anomaly ($E=0$) this
{\it stationary} generating function, $\Phi_{an}(u,\phi)$, obeys the
partial differential equation (PDE) \cite{KY-PRB, KY-Ann}
\begin{eqnarray}\label{anomal-equation}
&&\left[[1 - \cos{(4\phi)}]\,u^{2}\partial^{2}_{u} +
\sin{(4\phi)}\, u\partial_{u}\partial_{\phi}
+ \frac{3 +
\cos{(4\phi)}}{4}\partial^{2}_{\phi}
\right.
\nonumber \\
&&
\left.
+ 2\cos{(4\phi)}\,u\partial_{u}
- \frac{3}{2}\sin{(4\phi)}\partial_{\phi} - 2\cos{(4\phi)} -
u\right] \Phi_{an}(u,\phi) = 0
 \,
\end{eqnarray}
Its solution should also meet the requirements of being a smooth periodic
function of $\phi$, regular, positive and non-zero at $u\rightarrow
0$ (we recall that $\Phi(u=0, \phi)$ is the phase distribution
function, see Eq.(\ref{u=0})) and decaying at $u\rightarrow \infty$.

These requirements are rather restrictive. For instance, the
solution
\begin{equation}\label{SUSY-solution}
\Phi_{0}(u,\phi) = u\exp{\left(-\sqrt{u} \,\,(|\cos{\phi}| + |\sin{\phi}|)\right)}
\,
\end{equation}
is not appropriate for it is not a smooth function of $\phi$.

For comparison, we write down also the equation for the ``normal" generating
function $\Phi_{norm}(u,\phi)$ (i.e. when the energy lies outside the
anomaly region, or for the continuous model):
\begin{equation}
\label{ord}
\left[u^{2}\partial^{2}_{u}-u+\frac{3}{4}\partial^{2}_{\phi} \right]
\,\Phi_{norm}(u,\phi) = 0 \, .
\end{equation}
This equation looks like a ``course-grained" PDE
(\ref{anomal-equation}) where all the coefficients are ``averaged"
over the angle interval $(0,\pi)$ (so called ``phase randomization")
which is equivalent to course-graining over the space region
$\ell_{0}\gg\Delta x \gg 1/k$. The variables $u$ and $\phi$ in
Eq.(\ref{ord}) are separated and one immediately finds that the only
solution decaying at $u \rightarrow \infty$ and remaining regular
and non-zero at $u \rightarrow 0$ is given by
\begin{equation}\label{ord-solution}
 \Phi_{norm}(u,\phi) = \frac{2}{\pi}\sqrt{u}\,K_{1}(2\sqrt{u})
 \approx \frac{u^{1/4}}{\sqrt{\pi}}\,\mathrm{e}^{-2\sqrt{u}}  \,\,\,\,\, \mathrm{at}
\,\,\,\,\, u \gg 1 \, ,
\end{equation}
where $K_{1}(x)$ is the modified Bessel function.
This solution has been earlier obtained \cite{Kolok} in the continuous model.
It also arises in the theory of a multi-channel disordered wire
\cite{Efet-book,Mirlin2000}. The corresponding phase distribution is uniform:
$P_{norm}(\phi) = \Phi_{norm}(u=0,\phi) = 1/\pi$.

Unlike Eq.(\ref{ord}), the PDE (\ref{anomal-equation}) is not
separable in the variables $u$ and $\phi$. However, due to a hidden
(and not well established yet) symmetry of the problem, it was
possible to find new variables which allowed us to split the PDE
(\ref{anomal-equation}) into two ordinary differential equation and
thus to construct an exact general solution \cite{KY-PRB,KY-Ann}.
The solution, which  obeys the above requirements, is given by
\cite{KY-Ann}:
\begin{eqnarray}\label{canon-final}
\Phi_{an}(u, \phi)
&=&\frac{u^{1/2}}{2\Gamma^4\left(\frac{1}{4}
\right)|\cos{\phi}\,\sin{\phi}|^{1/2}} \int_{0}^{\infty}d\lambda\,
\frac{|\Gamma\left(\frac{1}{4}+\epsilon\lambda
\right)|^2}
{\lambda^{3/2}} \nonumber \\
&& \left[W_{-\lambda\epsilon,\frac{1}{4}}\,\left(
\frac{\bar{\epsilon}\, u\cos^2{\phi}}{4\lambda}\right)W_{-\lambda\bar{\epsilon},\frac{1}{4}}\,\left(
\frac{\epsilon \, u\sin^2{\phi}}{4\lambda}\right)+ c.c. \, \right] \, ,
\end{eqnarray}
where $\epsilon = \mathrm{e}^{i\pi/4}$, $\bar{\epsilon} = \mathrm{e}^{-i\pi/4}$ and $W_{\lambda, \mu}(z)$ is the Whittaker function (For the second index $\mu =1/4$ 
the Whittaker function can be expressed also in terms of the parabolic cylinder function,  see, e.g. \cite{GR}). In the limit $u \rightarrow 0$
the expression (\ref{canon-final}) reproduces the phase distribution function 
$P_{an}(\phi) = \Phi_{an}(u=0, \phi)$:
\begin{eqnarray}\label{P0-solution}
\mathcal{P}_{an}(\phi) = \frac{4\sqrt{\pi}}{\Gamma^2(\frac{1}{4})} \, \frac{1}{\sqrt{3 + \cos{(4\phi)}}}
     \, ,
\end{eqnarray}
which was derived earlier \cite{Derrida} in a different way. It
shows that the phase distribution becomes non-uniform at the
center-of-band anomaly.

Our current task is to derive an asymptotic expression for
$\Phi_{an}(u, \phi)$ in the limit of large $u \gg 1$. The integrand
in Eq.(\ref{canon-final}) is too complicated for a brute force
attack. This is because both the arguments and the first indices of
the Whittaker functions are large (as is shown below, the leading
contribution to the integral comes from $\lambda \sim \sqrt{u}$) and
the standard \cite{GR} asymptotic expansions of these functions are
not applicable. Our approach will include three steps: first we will
represent Eq.(\ref{canon-final}) in the form which allows us to find
an asymptotic expression of the integrand; then we obtain the
asymptotic form Eq.(\ref{P-an-asymp}) of the generating function
$\Phi(u,\phi)$ at large $u$ (this asymptotic expression will be
obtained in the next subsection), and finally the ALS distribution
function $P(|\psi|^{2})$ will be found by a saddle-point integration
over $\phi$.

The generating function Eq.(\ref{canon-final}) is periodic in $\phi$
(with the period $\pi/2$) and symmetric with respect to the change
$\phi \rightarrow \pi/2 - \phi$. Therefore, it is sufficient to calculate
$\Phi(u,\phi)$ in the angular interval $(0, \pi/4]$. We exploit the following integral
representation of the Whittaker function
(cf. 9.222.1 \cite{GR}):
\begin{eqnarray}\label{W-GR-modified}
W_{-\lambda,\frac{1}{4}}\,(x) = \frac{\sqrt{2}\, x^{1/4}}{\Gamma(1/4 + \lambda)}
\int^{\infty}_{1}\mathrm{e}^{-xt/2}\left(\frac{t-1}{t+1}\right)^{\lambda}
\frac{d t}{(t^2 -1)^{3/4}}
\,
\end{eqnarray}
valid for $\Re x \geq 0$ and $\Re \lambda \geq 0$. Since the
integrand in Eq.(\ref{canon-final}) is an analytical function within
the sector $\pi/4 \leq \arg{\lambda} \leq \pi/4$, we rotate the
integration contour $\lambda \rightarrow \lambda
\mathrm{e}^{i\pi/4}$ and introduce a new integration
variable $z$ :
\begin{eqnarray}\label{z-variable}
\lambda = \frac{1}{4}\sqrt{\frac{u}{z}}
\, \, .
\end{eqnarray}
After these transformations, Eq.(\ref{canon-final}) takes the form:
\begin{eqnarray}\label{Phi-Wick-z}
\Phi(u,\phi) = \frac{2 \, \sqrt{u}}{\Gamma^4(1/4)} \Re \, \int^{\infty}_{0}
\frac{\mathrm{e}^{i\pi/4}\,  dz}{\sqrt{z}}\,
I_1(z, \phi) I_2(z, \phi)
\, .
\end{eqnarray}
Here
\begin{eqnarray}\label{I-definition}
I_{1(2)}(z, \phi) = \int^{\infty}_{1}
\frac{d t}{(t^2-1)^{3/4}}\exp{[-\sqrt{u}\, f_{1(2)}(t, z, \phi)]}
\, ,
\end{eqnarray}
where
\begin{eqnarray}\label{f1}
f_1(t, z, \phi) \equiv
\frac{1}{4\sqrt{z}}\ln{\left(\frac{t+1}{t-1}\right)} +
\frac{t\,\sqrt{z}\, \cos^2{\phi}}{2}
\,
\end{eqnarray}
is real, while
\begin{eqnarray}\label{f2}
f_2(t, z, \phi) \equiv
-\frac{i}{4\sqrt{z}}\ln{\left(\frac{t+1}{t-1}\right)} +
\frac{i\, t\,\sqrt{z}\,\sin^2{\phi}}{2}
\,
\end{eqnarray}
is purely imaginary for real $z$.
Exact Eqs.(\ref{Phi-Wick-z})-(\ref{f2}) constitute the
starting point for the calculation of asymptotic
expressions at $u \gg 1$.

\subsection{Asymptotic of the integrand in Eq.(\ref{Phi-Wick-z})}
\begin{figure}[h]
\includegraphics[height=12cm,width=16cm]{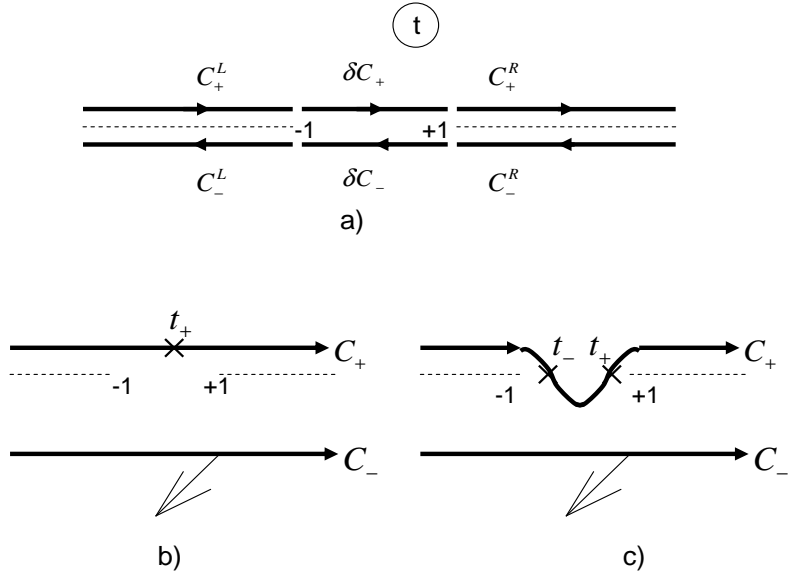}
\caption{Choice of contours (solid lines with arrows) in the complex
plane of $t$: a) initial contours; b) and c) the final contours
deformed to pass through the saddle points. The cuts are denoted by
the dotted lines. The contour $C_{-}$ can be deformed away to
infinity in the lower half-plane.} \label{cont-t}
\end{figure}
At $u\gg 1$ the integrals Eq.(\ref{I-definition}) can be computed in
 the saddle-point approximation. The minimum of the action in the
integrand of $I_{1}(z, \phi)$ is achieved at the point
\begin{eqnarray}\label{t0}
t_{0} = \sqrt{1 + \frac{1}{z\, \cos^2{\phi}}} > 1
\, .
\end{eqnarray}
The integration contour goes through this point, so the corresponding
saddle-point contribution is given by
\begin{eqnarray}\label{I-1-SP}
I^{(s)}_{1}(z, \phi) = \frac{\sqrt{2\pi}}{u^{1/4}}
\frac{z^{1/4}}{(1 + z\,\cos^2{\phi})^{1/4}}
\mathrm{e}^{-\sqrt{u} \, F_{1}(z, \phi)}
\, ,
\end{eqnarray}
where
\begin{eqnarray}\label{I1-saddle}
\hspace{-2.cm}
F_{1}(z, \phi) = f_1(t_{0}, z, \phi) =
\frac{\cos{\phi}}{2}
\left[ \frac{\ln{\left(\sqrt{1 + z \cos^2{\phi}}
+ \sqrt{z}\, \cos{\phi} \right)}}
{\sqrt{z}\, \cos{\phi}}
+ \sqrt{1 + z \cos^2{\phi}}\right]
\, .
\end{eqnarray}
For the integral $I_{2}(z, \phi)$ the situation is more complicated
as there are two saddle-points:
\begin{eqnarray}
t_{\pm} =
\pm \left\{  \begin{array}{cc}
    i\sqrt{\frac{1}{z\, \sin^2{\phi}} - 1} \, , & z < 1/\sin^2{\phi} \, , \label{t-small-z}\\
    \sqrt{1 - \frac{1}{z\, \sin^2{\phi}}} \, ,  & z > 1/\sin^2{\phi} \,\,  \label{t-large-z}\\
  \end{array}
\right. \, ,
\end{eqnarray}
both lie outside the integration semi-axis $t > 1$. On the complex
plane $t$ with two cuts, $(-\infty, -1)$ and $(1,\infty)$
we define an integral over a contour $C$ by
\begin{eqnarray}\label{I-C-def}
I_{2}[C] \equiv \int_{C}
\frac{d t}{(t^2-1)^{3/4}}\exp{[-\sqrt{u}\, f_{2}(t, z, \phi)]}
\, ,
\end{eqnarray}
where we choose the branch of the integrand so that on the upper
edge of the cut $(1, \infty)$
$I_{2}[C^R_{+}] = I_{2}(z, \phi)$. Taking the contour $C = C^{L}_{-}
+ C^{L}_{+}+ C^{R}_{-} + C^{R}_{+}$ as depicted in
Fig.\ref{cont-t}a, one checks straightforwardly that
\begin{eqnarray}\label{I-C}
I_{2}[C] &=& -2\mathrm{e}^{i\pi/4}
\mathrm{e}^{\pi\sqrt{u}/(2\sqrt{z})} \Re\left([1 +
i\mathrm{e}^{-\pi\sqrt{u}/(2\sqrt{z})}]\mathrm{e}^{i\pi/4} I_{2}(z,
\phi)\right)
\nonumber \\
&\approx& -2\mathrm{e}^{i\pi/4} \mathrm{e}^{\pi\sqrt{u}/(2\sqrt{z})}
\Re\left(\mathrm{e}^{i\pi/4} I_{2}(z, \phi)\right)\, .
\end{eqnarray}
Thus, with the exponential accuracy we have expressed the quantity
of our interest $\Re\left(\mathrm{e}^{i\pi/4} I_{2}(z, \phi)\right)$
(see Eq.(\ref{Phi-Wick-z})) in terms of the contour integral $I_{2}[C]$.
Evidently, the latter is not changed if the integration is extended
to parts $\delta C_{+}$ and $\delta C_{-}$ comprising the closed
contour (see Fig.\ref{cont-t}a). In this way we arrive at the
important relation:
\begin{eqnarray}\label{I-C-plus}
I_{2}[C] = I_{2}[C_{+}] + I_{2}[C_{-}] = I_{2}[C_{+}]
\, .
\end{eqnarray}
where the contours $C_{+}$ and $C_{-}$ are shown in
Fig.\ref{cont-t}b,c; the last equality in Eq.(\ref{I-C-plus}) holds
because the contour $C_{-}$ may be safely shifted down to the
infinitely remote part of the half-plane $\Im \, t < 0$, where
$I_{2}[C_{-}]$ vanishes (see Eqs.(\ref{I-C-def}) and (\ref{f2})).

Further transformations depend on the location of the saddle-points,
i.e. on the value of $z$, see
Eqs.(\ref{t-small-z}) and (\ref{t-large-z}). For $z \sin^2{\phi} < 1$,
the contour $C_{+}$ can be lifted to the upper half-plane of $t$
(see Fig.\ref{cont-t}b) to go through the saddle-point $t_{+}$
(\ref{t-small-z}) which provides the minimum of the action. The
corresponding saddle-point contribution to
$\Re\left(\mathrm{e}^{i\pi/4} I_{2}(z, \phi)\right)$ is given by
($z\sin^2{\phi} < 1$)
\begin{eqnarray}\label{I2-saddle-A}
\hspace{-2.4cm} \Re\left(\mathrm{e}^{i\pi/4} I^{(s)}_{2}(z,
\phi)\right) = -\frac{\mathrm{e}^{-i\pi/4}}{2}
\mathrm{e}^{-\pi\sqrt{u}/(2\sqrt{z})}I^{(s)}_{2}[C_{+}] =
\frac{\sqrt{2\pi}}{2u^{1/4}}\frac{ z^{1/4} \, \mathrm{e}^{-\sqrt{u} \,
 F^{<}_{2}(z,\phi)} }{[1 -
z\sin^2{\phi}]^{1/4}}
 \, ,
\end{eqnarray}
where
\begin{eqnarray}\label{F2-saddle-A}
\hspace{-2.6cm} F^{<}_{2}(z, \phi) = \frac{\sin{\phi}}{2} \left[
\frac{1}{\sqrt{z}\, \sin{\phi}} \left(\frac{\pi}{2} +
\arctan{\sqrt{\frac{1}{z \sin^2{\phi}} - 1}} \, \right) - \sqrt{1 -
z \sin^2{\phi}}\right].
\end{eqnarray}
When $z \sin^2\phi > 1$, the two saddle points (\ref{t-large-z}) lie
on the real axis. We bent the contour $C_{+}$ so that it goes
through the both points within the proper Stokes sectors
(Fig.\ref{cont-t}c). The resulting saddle-point contribution to
$\Re\left(\mathrm{e}^{i\pi/4} I_{2}(z, \phi)\right)$ is given by
\begin{eqnarray}\label{I2-saddle-B}
\hspace{-2.3 cm}
\Re\left(\mathrm{e}^{i\pi/4} I^{(s)}_{2}(z, \phi)\right) = \frac{\sqrt{2\pi}}{2u^{1/4}}
\frac{ z^{1/4}}{[z\sin^2{\phi} - 1]^{1/4}}
\left[\mathrm{e}^{i\pi/4} \mathrm{e}^{-\sqrt{u} \, F^{>}_{2}(z,\phi)} + c.c.
\right] \,;\,\,\,\,\, z > \frac{1}{\sin^2{\phi}}
\, ,
\end{eqnarray}
where
\begin{eqnarray}\label{F2-saddle-B}
\hspace{-2.5cm}
F^{>}_{2}(z, \phi) = \frac{\sin{\phi}}{2}
\left[ \frac{-i}{\sqrt{z}\, \sin{\phi}} \left(
\ln{(\sqrt{z \sin^2{\phi} -1} + \sqrt{z} \sin{\phi})} +
\frac{i\pi}{2}\right) + i\sqrt{z \sin^2{\phi} -1}\right] .
\end{eqnarray}
The two saddle-point expressions for $\Re\left(\mathrm{e}^{i\pi/4} I_{2}(z, \phi)\right)$,
Eqs.(\ref{I2-saddle-A}) and (\ref{I2-saddle-B}), can be represented by a single formula valid for
an arbitrary $z>0$:
\begin{eqnarray}\label{I2-saddle}
\hspace{-2.3 cm}
\Re\left(\mathrm{e}^{i\pi/4} I^{(s)}_{2}(z, \phi)\right) = \frac{\sqrt{2\pi}\, z^{1/4}}{2 \, u^{1/4}}
\Re \left\{\frac{\mathrm{e}^{i\pi/4} \mathrm{e}^{-\sqrt{u} \, F^{(+)}_{2}(z,\phi)} +
\mathrm{e}^{-i\pi/4} \mathrm{e}^{-\sqrt{u} \, F^{(-)}_{2}(z,\phi)}}
{[z\sin^2{\phi} - 1]^{1/4}}\right\}
\, ,
\end{eqnarray}
where
\begin{eqnarray}\label{F2-saddle}
\hspace{-1.5cm}
F^{(\pm)}_{2}(z, \phi) = \mp i\,\frac{\sin{\phi}}{2}
\left[ \frac{\ln{\left(\pm i\sqrt{z \sin^2{\phi} -1} \pm i\sqrt{z}\,\sin{\phi}\right)}}
{\sqrt{z}\, \sin{\phi}}
- \sqrt{z \sin^2{\phi} -1} \, \, \right]  \, .
\end{eqnarray}
Eqs. (\ref{I2-saddle}) and (\ref{F2-saddle})
are defined on the complex plane $z$ with a cut along the ray
$(1/\sin^2{\phi}, \infty)$; branches of $(z \sin^2{\phi} -1)^{1/2}$ and
$(z \sin^2{\phi} -1)^{1/4}$ are chosen to be positive
on the upper edge of the cut, the (standard) branch of $\ln{w}$ is defined by the
requirement $\Im(\ln{w})= 0$ at $w>0$ and the cut along $(-\infty,0)$ on the $w$-plane.
Accounting for Eqs. (\ref{I1-saddle}) and (\ref{I2-saddle}), we arrive at the
expression for the generating function Eq.(\ref{Phi-Wick-z}) in the form
($u \gg 1$):
\begin{eqnarray}\label{Phi-Wick-z-asymp}
\hspace{-1.5cm}
&& \Phi^s_{an}(u,\phi) = \frac{2\pi}{\Gamma^4(1/4)} \Re \, \int_{C} dz \,
\frac{\mathrm{e}^{i\pi/4} \mathrm{e}^{-\sqrt{u} \, \mathcal{F}_{+}(z,\phi)} +
\mathrm{e}^{-i\pi/4} \mathrm{e}^{-\sqrt{u} \,
\mathcal{F}_{-}(z,\phi)}}{[(z\,\cos^2{\phi} +1)(z\sin^2{\phi} -1)]^{1/4}} \, ; \\
&& \mathcal{F}_{\pm}(z,\phi) \equiv F_{1}(z, \phi) + F^{(\pm)}_{2}(z, \phi)
\label{F-cal}
\, .
\end{eqnarray}
\begin{figure}[h]
\includegraphics[height=6cm,width=8cm]{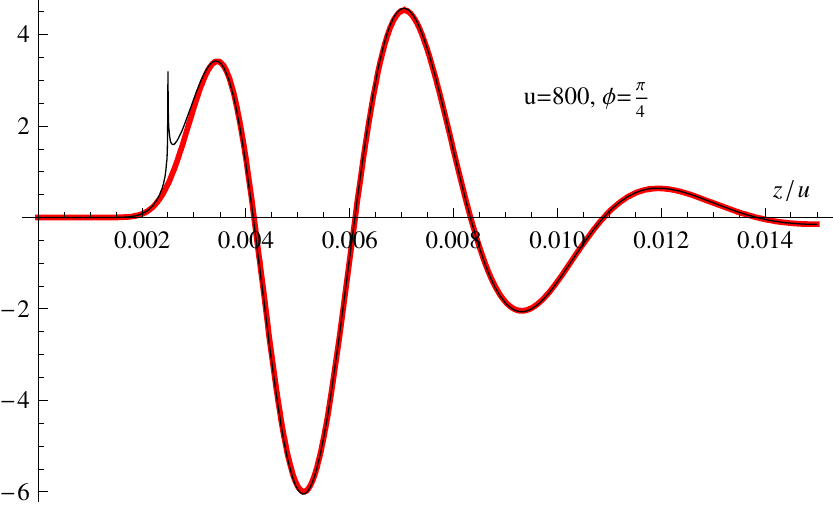}
\includegraphics[height=6cm, width=8cm]{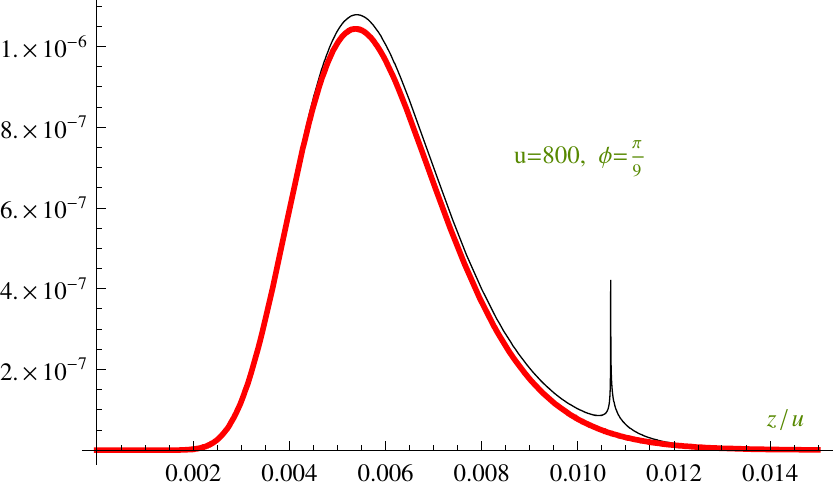}
\caption{The saddle-point integrand (arbitrary units) in
Eq.(\ref{Phi-Wick-z-asymp}) for $u=800$ and two different angles
$\phi=\frac{\pi}{4}$ and $\phi=\frac{\pi}{9}$ (thin solid line)
compared with the proper integrand in the exact solution
Eq.(\ref{canon-final}) (thick solid line). The coincidence is very
good except in the vicinity of the branch cut point
$z=\sin^{-2}\phi$ where there is an integrable singularity in the
saddle-point integrand. As $u$ increases, this peak singularity
moves to the
tails of the integrand (to the right tail for $\phi<\phi_{c}$ and to
the left tail for $\phi>\phi_{c}$) and thus makes  negligible
contribution to the $z$-integral. An exception is the case of
$\phi\approx\phi_{c}$ where the peak does not move to the tails. In
this case the saddle-point integrand Eq.(\ref{Phi-Wick-z-asymp}) is
no longer valid (see Appendix A).} \label{match-integrand}
\end{figure}
Here the integration contour $C = C_0 + C^R_{+}$ on the complex
plane $z$ with the cuts along the rays $(-\infty, -1/\cos^2{\phi})$
and $(1/\sin^2{\phi}, \infty)$, is shown in Fig.\ref{cont-z}; the
chosen branch of $(z\,\cos^2{\phi} +1)^{1/4}$ is positive at $z >
-1/\cos^2{\phi}$. The integrand in Eq.(\ref{Phi-Wick-z-asymp}) is
depicted (for different values of $\phi$) in
Fig.{\ref{match-integrand}} together with the result of the direct
numerical evaluation of the integrand (after switching to the
$z$-variable) in Eq.(\ref{canon-final}).

Our next step is the calculation of the integral in
Eq.(\ref{Phi-Wick-z-asymp}).

\subsection{Saddle-point calculation of the integral in Eq.(\ref{Phi-Wick-z-asymp})
for $\Phi(u,\phi)$}
\begin{figure}[h]
\includegraphics[height=7cm,width=11cm]{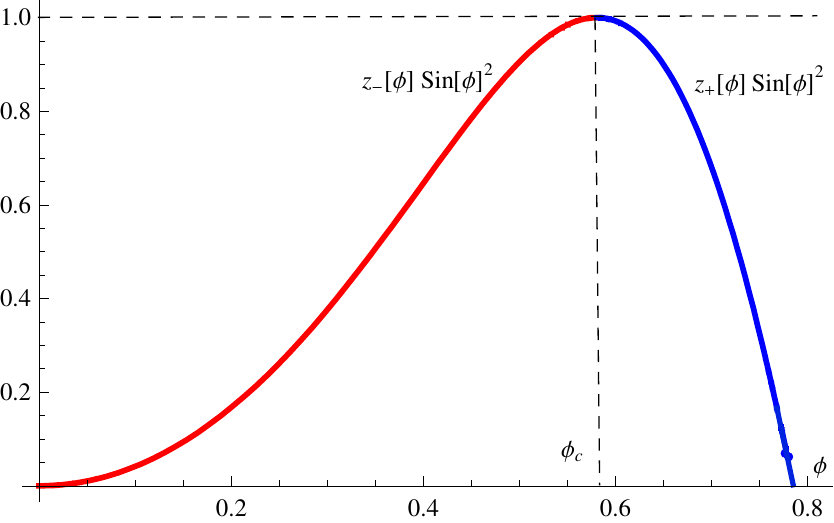}
\caption{ Solutions $z_{\pm}(\phi)$ to the saddle-point equations
Eq.(\ref{F-cal-saddle})} \label{z-1-2}
\end{figure}
\begin{figure}[h]
\includegraphics[height=12cm,width=16cm]{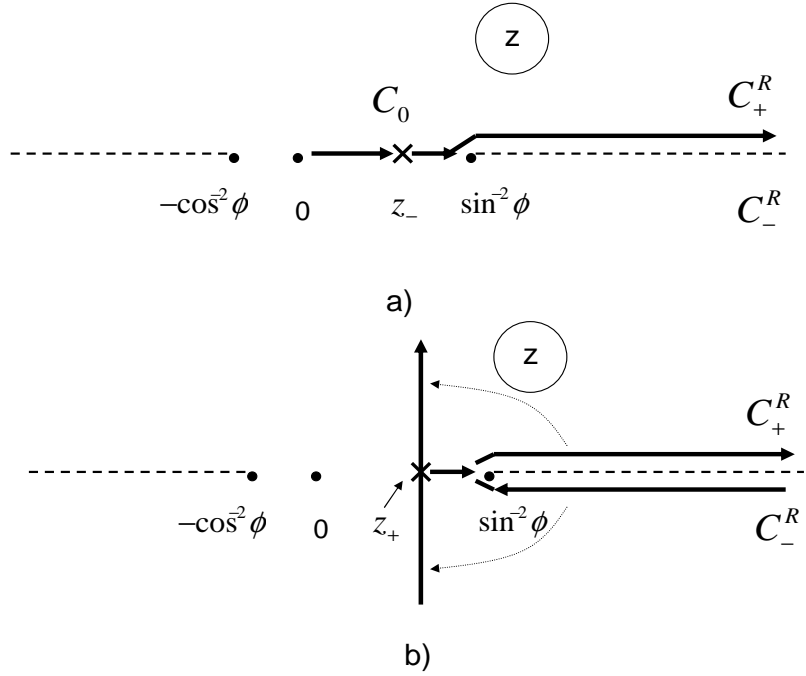}
\caption{Contours in the complex plane of $z$: a) evaluation of the
contribution of the saddle-point $z_{-}$; b) evaluation of the
contribution of the saddle-point $z_{+}$. At the critical angle
$\phi_{c}$ the saddle points (denoted by a cross) touch the branch
cut point $z=\sin^{-2}\phi$.} \label{cont-z}
\end{figure}

Saddle-points of the integrand in Eq.(\ref{Phi-Wick-z-asymp}) are
determined by solutions $z_{\pm}(\phi)$ to the equations
\begin{eqnarray}\label{F-cal-saddle}
\hspace{-1.5cm}&& \frac{\partial \mathcal{F}_{\pm}(z,
\phi)}{\partial z} =\frac{1}{4 z} \left[
-\frac{\ln{\left(\sqrt{z\cos^2{\phi} + 1} +
\sqrt{z}\cos{\phi}\right)}}{\sqrt{z}}
+ \cos{\phi} \sqrt{z\cos^2{\phi} + 1}  \right. \nonumber \\
&& \left. \pm i \frac{\ln{\left(\pm i\sqrt{z\sin^2{\phi} - 1} \pm i\sqrt{z}\sin{\phi}\right)}}{\sqrt{z}}
\pm i\sin{\phi} \sqrt{z\sin^2{\phi} - 1}
\right] = 0
 \, .
\end{eqnarray}
It turns out that the solutions $z_{\pm}(\phi)$ are real and $0 \leq
z_{\pm}(\phi) \leq 1/\sin^2{\phi}$; the solution $z_{-}(\phi)$
exists for $0 \leq \phi \leq \phi_{c}$, while the solution
$z_{+}(\phi)$ exists for $\phi_{c} \leq \phi \leq \pi/4$ (we recall
that we consider the angle interval $(0, \pi/4)$), see
Fig.\ref{cont-z}. The critical angle $\phi_{c}$ is determined by the
condition $z_{\pm}(\phi_{c})\sin^2{\phi_{c}} = 1$, i.e. the solution
reaches the origin of the right cut. With this condition, the
equation Eq.(\ref{F-cal-saddle}) results in (for the both signs):
\begin{eqnarray}\label{crit-angle}
\hspace{-1.5cm}
Y(\phi_c) = 0 \,\, , \,\,\,\,\,\, \mathrm{where} \,\,\,\, Y(\phi) \equiv
\ln{\left(\frac{\sin{\phi}}{1 + \cos{\phi}}\right)} +
\frac{\cos{\phi}}{\sin^2{\phi}} - \frac{\pi}{2}
\, .
\end{eqnarray}
This transcendental equation can be represented in a nice form using
the parametrization
\begin{eqnarray}\label{crit-angle-parametrization}
\hspace{-1.5cm}
\cot{\phi_c} = \sinh\left(\frac{x_c}{2}\right) \, ,
\end{eqnarray}
where $x_c \approx 2.4164...$ is the solution of the equation
\begin{eqnarray}\label{crit-angle-nice}
\hspace{-1.5cm}
\sinh{x} - x = \pi \,
\end{eqnarray}
The value of the critical angle $\phi_c$
\begin{eqnarray}\label{crit-angle-number}
\hspace{-1.5cm} \phi_c = 0.58060... \,
\end{eqnarray}
arises as an important constant also in the calculation of the
probability function $P_{an}(|\psi|^2)$, section \ref{Asymptotic of P}.

In the vicinity of this critical angle we have:
\begin{eqnarray}\label{z-crit}
z_{+}(\phi)&=&z_{-}(\phi)=\frac{1}{\sin^{2}(\phi)}\,\left(1-\frac{Y^{2}(\phi)}{4}\right)\\
\nonumber &\approx&
\frac{1}{\sin^{2}(\phi_{c})}-\frac{(\phi-\phi_{c})^{2}}{\sin^{4}\phi_{c}}\,\left(
\frac{1}{\sin^{2}\phi_{c}}-1\right),
\end{eqnarray}
where $Y(\phi)$ is given by Eq.(\ref{crit-angle}).
 At the ends of the angle interval, i.e. at $\phi =
\pi/4$ and $\phi = 0$, the solutions to Eq.(\ref{F-cal-saddle}) are
given by
\begin{eqnarray}
&& z_{+}(\phi) \approx 20\,\left( \frac{\pi}{4}-\phi\right)
-\frac{1000}{9}\,\left( \frac{\pi}{4}-\phi\right)^{3} \label{z-phi-max} \\
&& z_{-}(0) = \sinh^2(x_0/2) \approx 4.1263... \,\, ; \,\,
\mathrm{where} \,\,
 \sinh(x_0) - x_0 = 2\pi \, . \label{z-phi-min}
\end{eqnarray}
Using Eq.(\ref{F-cal-saddle}), one can represent  the saddle-point
actions in a following compact form:
\begin{eqnarray}\label{F-cal-saddle-action}
\hspace{-1.5cm} \mathcal{F}^{s}_{\pm}(\phi) = \cos{\phi} \, \sqrt{1
+ z_{\pm}(\phi)\cos^2{\phi}} \pm \sin{\phi} \, \sqrt{1 -
z_{\pm}(\phi)\sin^2{\phi}}
 \, .
\end{eqnarray}
In a similar way, the second derivatives of the actions in the saddle point can be
represented as:
\begin{eqnarray}\label{F-cal-saddle-2nd}
\hspace{-1.5cm} \left. \frac{\partial^2\mathcal{F}^{s}_{\pm}(z,
\phi)}{\partial z^2}\right|_{z=z_{\pm}(\phi)} = \frac{1}{4
z_{\pm}(\phi)} \left( \frac{\cos^3{\phi}}{\sqrt{1 +
z_{\pm}(\phi)\cos^2{\phi}}} \mp \frac{\sin^3{\phi}}{\sqrt{1 -
z_{\pm}(\phi)\sin^2{\phi}}} \right) \, .
\end{eqnarray}
It follows from Eq.(\ref{F-cal-saddle-2nd}) that the second
derivative is positive at $z_{-}(\phi)$ ($\phi < \phi_c$) and
negative at $z_{+}(\phi)$ ($\phi >\phi_c$). Therefore, for $\phi <
\phi_c$, the contour $C = C_0 + C^R_{+}$ (see Fig.\ref{cont-z}a)
goes through the saddle point $z_{-}(\phi)$ within the proper Stokes
sectors, and the term with ${\cal F}_{-}$ makes the contribution to
the integral in Eq.(\ref{Phi-Wick-z-asymp}). The term with ${\cal
F}_{+}$ in Eq.(\ref{Phi-Wick-z-asymp}) does not have a saddle point
at $\phi < \phi_{c}$ and its contribution is negligible at large
$u$.

\begin{figure}[h]
\includegraphics[height=7cm,width=11cm]{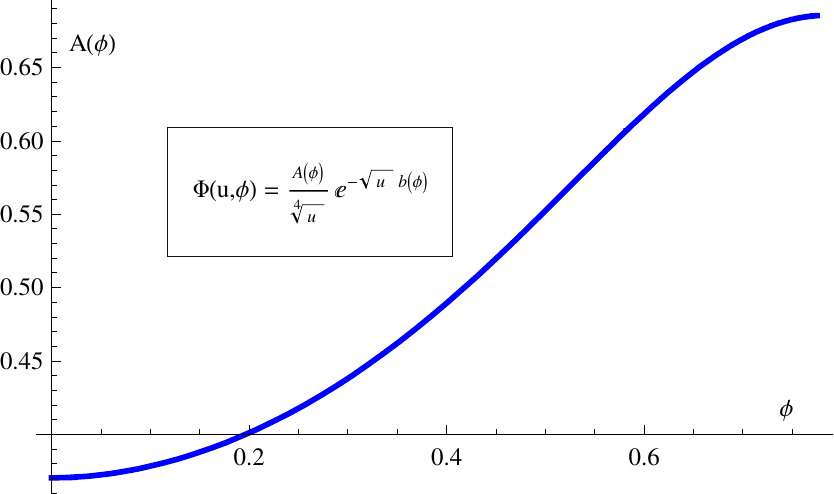}
\caption{The function $A(\phi)$ in
Eq.(\ref{Phi-Wick-z-asymp-general}). } \label{A-phi}
\end{figure}

\begin{figure}[h]
\includegraphics[height=7cm,width=11cm]{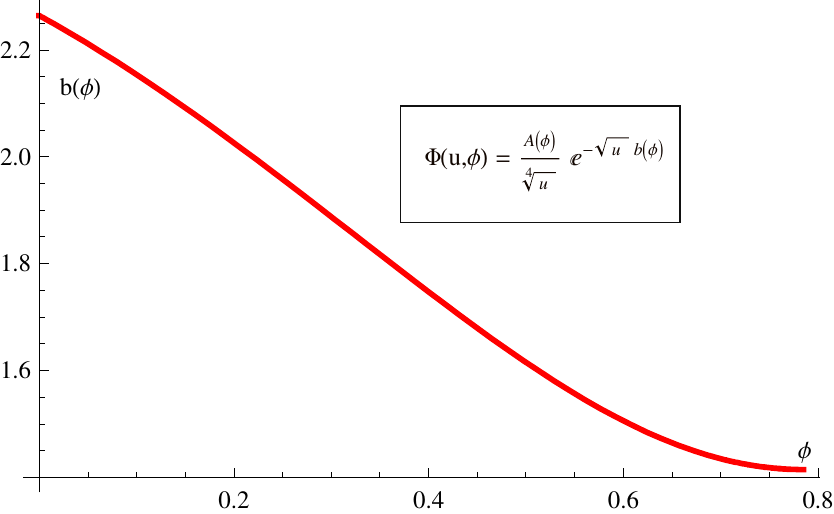}
\caption{The function $b(\phi)$ in
Eq.(\ref{Phi-Wick-z-asymp-general}). } \label{b-phi}
\end{figure}

On the contrary, for $\phi >\phi_c$ the contour $C$ is not
appropriate as it goes within improper Stokes sectors of the saddle
point $z_{+}(\phi)$. To overcome this obstacle, let us modify the
contour $C = C_{0} + C^R_{+}$ by adding an additional contour
$C^R_{-}$ which corresponds to the lower edge of the cut, see
Fig.\ref{cont-z}b. It is seen easily that this operation does not
change the integral Eq.(\ref{Phi-Wick-z-asymp}) because its
integrand is purely imaginary on $C^R_{-}$. Now, the part $C^R
\equiv C^R_{+} + C^R_{-}$ of the modified contour can be deformed to
the vertical contour which goes through the saddle point
$z_{+}(\phi)$ within the proper Stokes sectors, Fig.\ref{cont-z}b.
The corresponding saddle-point contribution to the generating
function Eq.(\ref{Phi-Wick-z-asymp}) at $\phi > \phi_c$ is
determined by the $\mathcal{F}_{+}(z,\phi)$ term in the integrand.
Summarizing these results we arrive at the following asymptotic
expression for the generating function:
\begin{eqnarray}\label{Phi-Wick-z-asymp-general}
\hspace{-1.5cm}
\Phi^{s}_{an}(u,\phi) =  A(\phi)
\frac{\mathrm{e}^{-\sqrt{u} \, b(\phi)}}{u^{1/4}} \, \,\,\,\,; \,\,\, u \gg 1 \, .
\end{eqnarray}
Here the function $A(\phi)$ outside of a narrow vicinity of the critical angle $\phi_c$
(see below) is given by
\begin{eqnarray}\label{AA-phi}
\hspace{-1.5cm} A(\phi) = \frac{(2\pi)^{3/2}}{\Gamma^4(1/4)}
\frac{2\sqrt{z_{\pm}(\phi)}} {\left|\, \sin^3{\phi} \, \sqrt{1+
z_{\pm}(\phi)\cos^2{\phi}} \mp \cos^3{\phi} \,
\sqrt{1-z_{\pm}(\phi)\sin^2{\phi}} \, \right|^{1/2}} \, ,
\end{eqnarray}
while the function $b(\phi)$ is given by
Eq.(\ref{F-cal-saddle-action}):
\begin{eqnarray}\label{bb-phi}
\hspace{-1.5cm}
b(\phi) = \mathcal{F}^{s}_{\pm}(\phi)  \, .
\end{eqnarray}
In these equations the upper (lower) signs stand for $\phi > \phi_c$
($\phi < \phi_c$).  At the particular angles, $\phi=\phi_{c}$, $\phi
=\pi/4$ and $\phi = 0$, the functions $b(\phi)$ and $A(\phi)$ are
given by (see expressions Eqs.(\ref{z-phi-max}) and
(\ref{z-phi-min})):
\begin{eqnarray}\label{b-phi-particular}
\hspace{-1.5cm} &&b(\phi = \pi/4) = \sqrt{2} \,\,\,\,;
\,\,\,\,\,\,\,\,\, b(\phi=0) =\sqrt{1+z_{-}(0)} \approx 2.2641... \\
\nonumber
&&b(\phi = \phi_{c} + \delta\phi)=\cot\phi_{c}- \delta\phi + \cot^{3}\phi_{c}\,\frac{(\delta\phi)^{2}}{2}+
O((\delta\phi)^{3}).
\end{eqnarray}
\begin{eqnarray}\label{A-phi-particular}
\hspace{-1.5cm} &&A(\phi = \pi/4) = \frac{8\,
2^{\frac{1}{4}}\,\sqrt{5}\,\pi^{\frac{3}{2}}}{\Gamma^{4}(\frac{1}{4})}=
0.6855... \,\, ,\\ &&A(\phi=0)
=\frac{2(2\pi)^{\frac{3}{2}}}{\Gamma^{4}(\frac{1}{4})}\,\sqrt{z_{-}(0)}
\approx 0.3703 \\  &&A(\phi=\phi_{c})=\frac{4\,
\sqrt{2}\,\pi^{\frac{3}{2}}}{\Gamma^{4}(\frac{1}{4})\,\sin^{2}\phi_{c}}=
0.6059... \,\, .
\end{eqnarray}
The plots of the functions $A(\phi)$ and $b(\phi)$ computed from
Eqs.(\ref{AA-phi}) and (\ref{bb-phi}) are given in Fig.\ref{A-phi}
and Fig.\ref{b-phi}. Remarkably, the plots which were calculated
from different expressions at $\phi>\phi_{c}$ and $\phi<\phi_{c}$ do
not show any singularity at $\phi=\phi_{c}$. The two pieces of the
curves match perfectly at the critical angle $\phi=\phi_{c}$.

In the next section we present a different calculation of the
function $b(\phi)$ which does not possess any
critical angle by construction and coincides identically with
the above saddle-point expressions. As both $A(\phi)$ and $b(\phi)$
are expressed through the same solutions $z_{\pm}(\phi)$ of the
saddle-point equation, smoothness of $b(\phi)$ at $\phi=\phi_{c}$
implies also the smoothness of $A(\phi)$.
\begin{figure}[h]
\includegraphics[height=7cm,width=11cm]{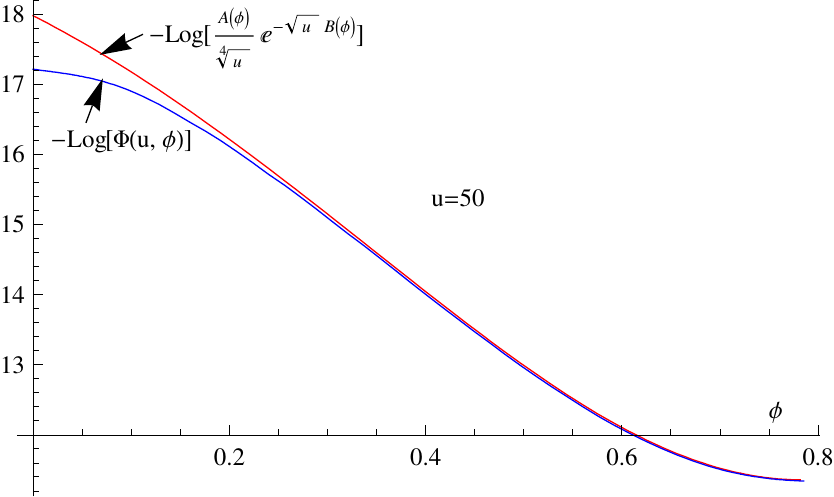}
\caption{The comparison of the approximate function $\Phi(u,\phi)$
given by Eq.(\ref{Phi-Wick-z-asymp-general}) and the exact
$\Phi(u,\phi)$ Eq.(\ref{canon-final})obtained numerically at $u=50$.
Near $\phi=0$ the exact function has an extremum to ensure
smoothness of the even function $\Phi(u,\phi)=\Phi(u,-\phi)$. The
approximation is not accurate in the vicinity of $\phi=0$, where the
right branch-cut point in Fig.\ref{cont-z} moves to infinity.
However, it becomes more and more accurate as $u$ increases. Note
that the principle parameter of the approximation $u^{-1/4}\approx
0.38$ is not very small at $u=50$. } \label{Log-Phi-compar}
\end{figure}

\subsection{Ordinary differential equation for the exponent $b(\phi)$}

Let us look for an asymptotic ($u \gg 1$) solution to the original
PDE (\ref{anomal-equation}) in the form:
$\Phi(u, \phi) \sim \exp{[-u^p b(\phi)]}$ where $m$ and $b(\phi)$ are
to be determined by keeping in the PDE terms of the leading order in $u$.
We find immediately that $p=1/2$ (which is in accordance with
(\ref{Phi-Wick-z-asymp-general})) while $b(\phi)$ obeys the ordinary
differential equation (ODE):
\begin{eqnarray}\label{Diff-Eq-for-b}
\hspace{-1.5cm}
\frac{3 + \cos{4\phi}}{4} \left(\frac{d b(\phi)}{d \phi}\right)^2 +
\frac{\sin{4\phi}}{2} b(\phi) \frac{d b(\phi)}{d \phi} +
\frac{1 - \cos{4\phi}}{4}\, b^2(\phi) = 1
\, .
\end{eqnarray}
One can reduce the equation to the form convenient for numerical
integration by introducing the function:
\begin{equation}
\label{y}
y(\phi)=\frac{\sqrt{2}\,b(\phi/2)}{2(1+\cos^{2}\phi)^{\frac{1}{4}}}.
\end{equation}
Then Eq.(\ref{Diff-Eq-for-b}) takes the form:
\begin{equation}
\label{ODE-can} \frac{dy}{d\phi}=\pm
\frac{\sqrt{1-y^{2}\,\frac{\sin^{2}\phi}{(1+\cos^{2}\phi)^{1/2}}}
}{2(1+\cos^{2}\phi)^{3/4}}.
\end{equation}
The initial conditions for Eqs.(\ref{Diff-Eq-for-b}),(\ref{ODE-can})
follow from Eq.(\ref{b-phi-particular}):
\begin{equation}\label{init}
b(\pi/4)=\sqrt{2},\;\;\;\;y(\pi/2)= 1.
\end{equation}
There is an obvious solution to Eq.(\ref{Diff-Eq-for-b}) with the
initial condition Eq.(\ref{init}):
\begin{equation}
\label{obv} b_{0}(\phi)=\cos\phi+\sin\phi.
\end{equation}
It corresponds to the choice of sign "+" in Eq.(\ref{ODE-can}). This
solution is a growing function of $\phi$ with the {\it maximum} at
$\phi=\pi/4$. Therefore it does not correspond to the saddle-point
solution which has a {\it minimum} at $\phi=\pi/4$ (see
Fig.\ref{b-phi} and Eq.(\ref{b-phi-particular})). In fact, the
solution Eq.(\ref{obv}) corresponds to the particular solution
Eq.(\ref{SUSY-solution}) which we have already discarded on physical
grounds. Thus the relevant solution for our problem is the one which
corresponds to the sign "minus" in Eq.(\ref{ODE-can}). This ODE can
be transformed into the Abel's ODE \cite{Kamke} but it does not
belong to the classes with known solutions.

We solved Eq.(\ref{ODE-can}) numerically applying the initial
condition Eq.(\ref{init}) at a point $\phi=\pi/2-\delta$ with
$\delta=10^{-10}$. We checked that the solution corresponding to the
sign "plus" matches the function $b_{0}(\phi)$ obtained from
Eqs.(\ref{y}),(\ref{obv}) with the same accuracy. Much less trivial
is that the solution for $b(\phi)$ corresponding to the sign "minus"
in Eq.(\ref{ODE-can}) coincides (with the same accuracy) with the
saddle-point solution given by
Eqs.(\ref{F-cal-saddle-action}),(\ref{bb-phi}). Remarkably, the
solution to the particular Abel's ODE appeared to be represented in
terms of the solution to the transcendental saddle-point equation
Eq.(\ref{F-cal-saddle})! In this connection we would like to remind
about another "miracle" of the problem. Namely, the saddle-point
solution for $b(\phi)$ which we obtained from two pieces ${\cal
F}^{s}_{\pm}(\phi)$ expressed through solutions $z_{+}(\phi)$ and
$z_{-}(\phi)$ of the saddle-point equations Eq.(\ref{F-cal-saddle}),
appeared to be smooth at the critical angle $\phi_{c}$ where the two
pieces match perfectly. Now we understand that this is a direct
consequence of the fact that $b(\phi)$ can be obtained from the ODE
which has no singularity at $\phi=\phi_{c}$.

This argument is also important to realize that the asymptotic
function $\Phi_{an}(u,\phi)$ (\ref{Phi-Wick-z-asymp-general}) is
valid also in the vicinity of the critical angle $\phi_{c}$ where
the saddle-point expression for the integrand in
Eq.(\ref{Phi-Wick-z-asymp}) is no longer valid. As is shown in
Appendix A, at $|\phi-\phi_{c}|<u^{-1/6}$ the integrand should
be modified so that the (fake) singularity at $z=\sin^{-2}\phi$
is rounded. Then
the $z$-integral can be computed analytically which results in the
same asymptotic Eq.(\ref{Phi-Wick-z-asymp-general}) with $b(\phi)$
given by Eq.(\ref{bb-phi}). We conclude therefore that
different procedures for $|\phi-\phi_{c}|\gg u^{-1/6}$ and
$|\phi-\phi_{c}|\ll u^{1/6}$ give the same result. Thus there is no
real ``critical angle" in the function $\Phi_{an}(u,\phi)$, while
there is a critical point in the integrand in
Eq.(\ref{Phi-Wick-z-asymp}).

\section{Probability distribution function $P_{an}(|\psi|^2)$ of anomalously localized eigenstates}
\label{Asymptotic of P}

The generating function $\Phi(u, \phi; E)$ allows one to calculate
all local statistics of eigenfunctions.
The probability distribution
function $P(|\psi|^2)$ is connected with a ``joint probability
distribution function" $P(u,\phi)$ (see \cite{KY-Ann} for details):
\begin{eqnarray}\label{P-psi-P-u-phi}
\hspace{-2.cm}
P(|\psi|^2) = \int^{\infty}_{0} d u \, \int^{\pi}_{0}
d\phi \, \delta(|\psi|^2 - u\cos^2{\phi}) P(u, \phi) = \int^{\pi}_{0}
\frac{d\phi}{\cos^2{\phi}}P\left(\frac{|\psi|^2}{\cos^2{\phi}}\, , \phi\right)\, .
\end{eqnarray}
The function $P(u,\phi)$, in its turn, is related with the
generating function $\Phi(u,\phi)$. This relation in the limit of a
long chain of the length $L\gg \ell_{0}$ reads:
\begin{eqnarray}\label{P-u-phi}
P(u,\phi)&=& i\frac{\nu_{0}(E)}{L\nu(E)
u}\,\partial_{u}\,\int_{-i\infty+0}^{+i\infty+0} \frac{dt}{t}\,
e^{4t/\ell_{0}}\, \Phi^2(u t, \phi)\\
\nonumber &=&-4i\,\frac{\nu_{0}(E)}{L\nu(E)
u^{2}}\,\int_{-i\infty+0}^{+i\infty+0} dt\, e^{4t/\ell_{0}}\,
\Phi^2(u t, \phi) \, ,
\end{eqnarray}
where the localization length $\ell_{0} \gg 1$ (away from the $E=0$
anomaly), the averaged DoS $\nu(E=0)$, and the DoS $\nu_{0}(E=0)$ of an
ideal (without disorder) chain are given by Eqs.(\ref{nu0-and-l0})
and (\ref{l-to-l0}).
Our aim is to find the asymptotic form of $P(|\psi|^2)$ at
$|\psi|^2\ell_{0} \gg 1 $.

The asymptotic form of the function $P(u,\phi)$ is determined by
that of the generating function $\Phi(u, \phi)$
\begin{eqnarray}\label{Phi-asymp-most-general}
\hspace{-1.5cm}
\Phi^{s}_{an}(u,\phi) =  \mathcal{A}(\phi) u^{q}
\mathrm{e}^{-\sqrt{u} \, b(\phi)}\, \,\,\,\,; \,\,\, u \gg 1
\,
\end{eqnarray}
represented in the form suitable for both the normal ($q = 1/4$, $b(\phi) = 2$; Eq.(\ref{ord-solution})) and anomalous ($q = -1/4$; Eq.(\ref{Phi-Wick-z-asymp-general}))
functions.

Plugging this function into Eq.(\ref{P-u-phi}) and doing the
saddle-point integration over $t$ one obtains:
\begin{equation}\label{Phi-u-phi}
\hspace{-1.5cm}
P(u,\phi)=2\sqrt{\pi}\,{\cal
A}^{2}(\phi)\,\frac{\nu_{0}(E=0)}{L\,\nu(E=0)}\,\left(
\frac{\ell_{0}}{4}\right)^{4q+1/2}\,[b(\phi)]^{4q+1}\,u^{4q-3/2}\,{\rm
e}^{-u\ell_{0}\,b^{2}(\phi)/4}.
\end{equation}
Now using the $\pi/2$-periodicity of the integrand in
Eq.(\ref{P-psi-P-u-phi}) one finally arrives at:
\begin{eqnarray}\label{P-psi-intermediate}
\hspace{-2.cm}
P(|\psi|^2)
=  C \, (|\psi|^2\ell_{0})^{4q-3/2}
\int^{\pi/2}_{0}\frac{\mathcal{A}^2(\phi) }{[\cos{\phi}]^{8q - 1}}
\; [b(\phi)]^{4q+1}\;
\mathrm{e}^{-\frac{|\psi|^2\ell_{0}}{4}\frac{b^2(\phi)}{\cos^2{\phi}}}\; d\phi \, ,
\end{eqnarray}
where
\begin{equation}
\label{C-E} C = \frac{2\sqrt{\pi}}{
4^{4q}}\,\frac{\ell^2_{0}}{L}\; \,\frac{\nu_{0}(E=0)}{\nu(E=0)}.
\end{equation}
In the limit $|\psi|^2\ell_{0} \gg 1$, the major contribution to
Eq.(\ref{P-psi-intermediate}) comes from the vicinity of the minimum
of the function
\begin{eqnarray}\label{B-phi-definition}
\hspace{-1.5cm}
B(\phi) \equiv \left(\frac{b(\phi)}{\cos{\phi}}\right)^2
\,
\end{eqnarray}
\begin{figure}
\includegraphics[height=7.5cm,width=10.5cm]{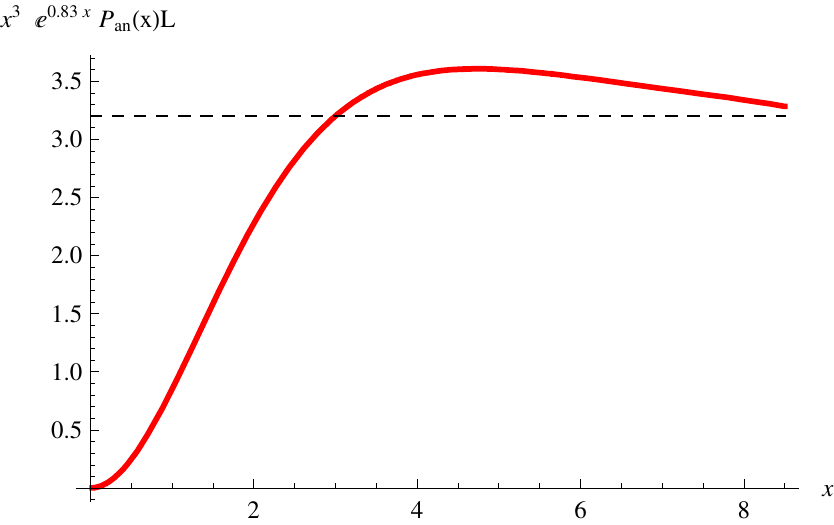}
\caption{The ratio of the probability distribution $P(x)$
 (computed numerically from the
exact solution Eq.(\ref{canon-final})) to the asymptotic expression
Eq.(\ref{P-psi-anomal-answer}) obtained for $x = |\psi|^{2}\ell_{0} \gg 1$.
The depletion at $x<4$
arises because of the $1/x$ behavior of the exact distribution at
small $x$ as compared to $1/x^{3}$ behavior of the asymptotic
expression. The quasi-constant behavior at $x>4$ (with the value of
the constant close to that of Eq.(\ref{C-phi-c}) depicted by the
dashed line) indicates on the setting up of the pre-exponent
$\propto 1/x^{3}$.} \label{P-x-compar}
\end{figure}
entering the exponent. Outside the anomaly (or for the continuum
model) the function $b(\phi) = 2$ (see Eq.(\ref{ord-solution}), so
the minimum value of $B(\phi)$ is achieved at $\phi = 0$. Performing
the saddle-point integration in Eq.(\ref{P-psi-intermediate}) we
obtain the announced expression Eq.(\ref{1d-dist}) for the
asymptotic of the ``normal" probability distribution function
$P_{norm}(|\psi|^2)$ of eigenstates. It is interesting that this
asymptotic form coincides with the exact function $P_{norm}(|\psi|^2)$
\cite{KY-Ann} for any value of $|\psi|^2$.

For the center-of-band anomaly, the function $b(\phi)$ is more
complicated. It follows from  Eq.(\ref{b-phi-particular}) that the
function $B(\phi)$ has its minimum exactly at the critical angle $\phi_c$
(Eqs.(\ref{crit-angle})-(\ref{crit-angle-number})):
\begin{eqnarray}\label{B-critical}
\hspace{-1.5cm} B(\phi_c + \delta \phi) = \frac{1}{\sin^2{(\phi_c)}}
+ \frac{(\delta \phi)^2}{\sin^4{(\phi_c)}}  + \ldots \, .
\end{eqnarray}
Thus the critical angle remarkably appears again in the theory even
though we have shown that the function $\Phi_{an}(u,\phi)$ is smooth
at $\phi=\phi_{c}$.

The saddle-point integration in Eq.(\ref{P-psi-intermediate}) leads
to the following result:
\begin{eqnarray}\label{P-psi-anomal-answer}
\hspace{-2.cm} P_{an}(|\psi|^2) =  C(\phi_{c})\,\frac{1} {\ell_{0}L\,|\psi|^{6}}
\exp{\left(-\frac{|\psi|^2\ell_{0}}{4\sin^2{(\phi_c)}} \right)} \, ,
\end{eqnarray}
where
\begin{equation}
\label{C-phi-c} C(\phi_{c})=\frac{64\pi\sqrt{2}}
{\Gamma^4\left(\frac{1}{4}\right)}  \,
\frac{\cos^{3}\phi_{c}}{\sin^{2}\phi_{c}}\approx 3.20.
\end{equation}
and
\begin{eqnarray}\label{P-exp}
\hspace{-1.5cm} \frac{1}{4\sin^2{(\phi_c)}} = 0.8310... < 1 \, ,
\end{eqnarray}
Equation (\ref{P-psi-anomal-answer}) is the main result of our
paper.
\section{$P(|\psi|^{2})$ at small $|\psi|^{2}\ell_{0}\ll 1$.}
In this section we consider the behavior of the eigenfunction
amplitude distribution function $P(|\psi|^{2})$ at small values of
$|\psi|^{2}\ell_{0}$. Generically, the small amplitudes
$|\psi|^{2}\ell_{0} \ll 1$ arise either
(i) due to localization when the observation
point ${\bf r}$ in $\psi=\psi({\bf r})$ lies outside the
localization volume, or (ii) due to the proximity of the observation
point to the node of the wave function. In the case (i) the
amplitude of exponentially localized eigenfunction cannot be smaller
than $|\psi|\sim  \ell^{-1/2}_{0} e^{-L/2\ell^{{\rm ext}}}$, while in the
case (ii) the amplitude $|\psi|$ can be arbitrary small. It is
clear that the case (i) is realized with almost certainty in a large
sample, while the case (ii) has small probability proportional to
the small distance of the observation point from the node. This
should lead to the drastically different behavior of the
distribution function for $|\psi|^{2}\ell_{0}\gg e^{-L/\ell^{{\rm
ext}}}$ (case (i)) and for $|\psi|^{2}\ell_{0}\ll e^{-L/\ell^{{\rm
ext}}}$ (case (ii)). Our approach based on the exact solution of
the {\it stationary}  (with respect to the coordinate along
the chain) evolution equation (\ref{anomal-equation}) is capable
of describing only the case (i), as the crossover to the alternative
case (ii) and the corresponding solution for the generating
function are essentially $L$-dependent.

Furthermore, one can argue that for the case of pure exponential
localization the asymptotic behavior of $P(|\psi|^{2})$ at small
$|\psi|^{2}\ell_{0}$
($e^{-L/\ell^{{\rm ext}}}\ll |\psi|^{2}\ell_{0}\ll 1$) should be
always $P(|\psi|^{2})=C_{{\rm norm}}/|\psi|^{2}$. Indeed, in this
case the normalization integral is logarithmically divergent and
dominated by $|\psi|^{2}\ell_{0}\sim e^{-L/\ell^{{\rm ext}}}$. Thus
the normalization constant $C_{{\rm norm}}\propto 1/L$, as it should
be in order to make the first moment $\langle
|\psi|^{2}\rangle =\frac{1}{L}$ as the eigenfunction normalization
requires. Should the profile of the localization tail be of the form
$L^{-\alpha}\,e^{-L/\ell^{{\rm ext}}}$ with an extra power-law
pre-exponent, the characteristic {\it sub-leading} terms  appear in
$P(|\psi|^{2})$:
\begin{equation}\label{small-asymp}
P(|\psi|^{2})=\frac{C_{{\rm norm}}}{|\psi|^{2}} (1-
\frac{\alpha}{\ln
|\psi|^{2}}+...),\;\;\;\;\;1\gg|\psi|^{2}\ell_{0}\gg e^{L/\ell^{{\rm
ext}}}.
\end{equation}
Thus studying details of the distribution function $P(|\psi|)$
at small amplitudes one may infer information about the profile of
the tail of the wave function.

In this section we briefly discuss how the principle term in
Eq.(\ref{small-asymp})arises in our formalism. To begin with we note
that according to Eq.(\ref{P-psi-P-u-phi}), the term
$|\psi|^{-2}$ at $|\psi|\ll 1$ may arise only when $P(u \ll 1, \phi) \propto
u^{-1} $. This means that the integral in Eq.(\ref{P-u-phi})
\begin{equation}\label{eqq}
\int_{-i\infty+0}^{+i\infty+0} \frac{dt}{t}\, e^{4t/\ell_{0}}\,
\Phi^2(u t, \phi)
\end{equation}
must be proportional to $u$ at $u\ll 1$. Then one immediately
concludes that the function $\Phi(u,\phi)$ should have a singularity
at $u=0$. Indeed, in case of a regular $u$-expansion, the term
$\propto u$ in $\Phi^{2}(u,\phi)$ would result in a linear in $u$
contribution in Eq.(\ref{eqq}) proportional to the integral:
$$
\int_{-i\infty+0}^{+i\infty+0} dt\, e^{4t/\ell_{0}}=0.
$$
To obtain the desired dependence $P(u \ll 1, \phi) \propto
u^{-1} $, one has to assume that there is
a term $\propto u \ln u$ in the expansion of $\Phi(u,\phi)$.
Then the corresponding integral in Eq.(\ref{eqq})
\begin{equation}\label{non-zero-int}
\int_{-i\infty+0}^{+i\infty+0} dt\,\ln t\, e^{4t/\ell_{0}}=-2\pi
i\int_{0}^{\infty}dt\, {\rm e}^{-4t/\ell_{0}}=-2\pi i
\,\frac{\ell_{0}}{4}.
\end{equation}
would be non-zero and result in $P(u,\phi)\propto u^{-1}$.
We see,
therefore, that the term $u\,\ln u$ in the expansion of
$\Phi(u,\phi)$ at small $u$ is the direct consequence of exponential
localization.

Next, one can check that that the ``normal" (away from the
anomaly) generation function  Eq.(\ref{ord-solution}) has,
indeed, the expansion with $\ln u$-terms:
\begin{equation}
\label{exp-ord}
\Phi_{norm}(u,\phi)=\frac{1}{\pi}\sum_{n=0}^{\infty}\left(
g_{n}+f_{n}\,\ln u\right)\,u^{n},
\end{equation}
where the first few coefficients of expansion are given by:
\begin{eqnarray}\label{part-val}
g_{0}&=&1,\;\;f_{0}=0,\;\;f_{1}=1,\;\;g_{1}= 2\gamma-1,\\
\nonumber f_{2}&=&\frac{1}{2},\;\; g_{2}=\gamma -\frac{5}{4},
\end{eqnarray}
where $\gamma = 0.577216...$ is the Euler constant.

A natural {\it assumption} would be that the generating function at
the anomaly $\Phi_{an}(u,\phi)$ has the same type of expansion
Eq.(\ref{exp-ord}) but with the $\phi$-dependent coefficients
$g_{n}(\phi)$ and $f_{n}(\phi)$:
\begin{equation}\label{exp-an}
\Phi_{an}(u,\phi)
\stackrel{\mathrm{?}}{=}
\sum_{n=0}^{\infty}\left[g_{n}(\phi)+f_{n}(\phi)\,\ln
u\right]\,u^{n}.
\end{equation}
If so, one can find the coefficients by plugging the series
Eq.(\ref{exp-an}) directly into Eq.(\ref{anomal-equation}). Then one
obtains the chain of recursive equations:
\begin{eqnarray}
\hspace{-1.5cm}
 \left[\frac{3 + \cos{4\phi}}{4}\partial^2_{\phi} + \left(n -
\frac{3}{2}\right)\sin{4\phi}\,\partial_{\phi}
+ n(n-1) - (n-1)(n-2)\cos{4\phi}\right]g_n(\phi)+ \nonumber \\
\hspace{-1.5cm}
 + \left[ \, \sin{4\phi}\,\partial_{\phi} + 2n-1 + (3-2n)\cos{4\phi}\right] f_n(\phi) =
 g_{n-1}(\phi)
     \label{f-n - equation} \\
     \hspace{-1.5cm}
\left[\frac{3 + \cos{4\phi}}{4}\partial^2_{\phi} + \left(n -
\frac{3}{2}\right)\sin{4\phi}\,\partial_{\phi} + n(n-1) -
(n-1)(n-2)\cos{4\phi} \, \right]f_n(\phi)= \nonumber
\\ \hspace{-1.5cm}
 = f_{n-1}(\phi) \label{g-n
- equation} \, .
\end{eqnarray}
For $n=1$ the equation (\ref{g-n - equation}) takes the form
\begin{eqnarray}\label{g-1 - equation}
\left[\frac{3 + \cos{4\phi}}{2}\partial_{\phi} -
\sin{4\phi}\right]\,\partial_{\phi} f_{1}(\phi) = 0 \, ,
\end{eqnarray}
which determines the derivative of $f_{1}(\phi)$:
\begin{eqnarray}\label{g-1 - first integral}
\partial_{\phi}f_{1}(\phi) = \frac{c_f}{\sqrt{3 + \cos{(4\phi)}}}.
\end{eqnarray}
Now we are going to apply the condition of periodicity of
$\Phi(u,\phi)$ as the function of the angle $\phi$. Since the
left-hand side of this equation is a derivative of a periodic
function, its integral over the period must vanish. This can be
provided only with the choice $c_f =0$. Hence, the function
$f_{1}(\phi)$ is a constant:
\begin{eqnarray}\label{g-1 - first integral}
f_{1}(\phi)=F_{1}\, ,
\end{eqnarray}
where the constant $F_{1}$ cannot be fixed by the homogeneous
equation (\ref{g-1 - equation}).

However, it appears that the requirement of periodicity of the
function $g_{1}(\phi)$ helps to fix the constant $F_{1}$. Indeed,
the equation for this function is:
\begin{eqnarray}\label{f-1 - equation}
\left[\frac{3 + \cos{4\phi}}{4}\partial_{\phi} -
\frac{1}{2}\sin{4\phi}\right]\,\partial_{\phi} g_{1}(\phi) =
g_{0}(\phi) - (1+ \cos{4\phi})F_1\, ,
\end{eqnarray}
where $g_{0}(\phi)\equiv{\cal P}_{an}(\phi)$ is given by
Eq.(\ref{P0-solution}). Looking for the solution in the form
$\partial_{\phi} g_{1}(\phi) = c_{g}(\phi)g_{0}(\phi)$ and taking
into account that $g_{0}(\phi)$ obeys the homogeneous equation (for
the zero R.H.S.), we obtain the following equation for
$c_{g}(\phi)$:
\begin{eqnarray}\label{c - equation}
\partial_{\phi} c_{g}(\phi) = \frac{4}{3 + \cos{4\phi}} -
\frac{\Gamma^2(\frac{1}{4})}{\sqrt{\pi}} \, \frac{1+
\cos{(4\phi)}}{\sqrt{3 + \cos{(4\phi)}}}F_1 \, .
\end{eqnarray}
Using once again the periodicity condition, we must require the
integral over the period of each side of the above equation to
vanish. This determines uniquely the value of $F_1$:
\begin{eqnarray}\label{F1}
F_{1}= \frac{\sqrt{2}}{4} \, .
\end{eqnarray}
We see that the solution can be found uniquely only if one assumes
the periodicity (and hence smoothness) of the function
$\Phi(u,\phi)$, and this solution corresponds to $f_{1}(\phi)\neq
0$. This means that it would not be  possible to find any periodic
solution without a term $\propto u\ln u$ in the series
Eq.(\ref{exp-an}) for $\Phi_{an}(u,\phi)$ . Thus
the periodicity requires the singular expansion at $u=0$ with
certainty. We note that the assumption of smoothness was the key
point to obtain the exact solution Eq.(\ref{canon-final})
\cite{KY-PRB, KY-Ann}.

With the coefficient $f_{1}(\phi)=F_{1}$ established one immediately
finds the leading term in the $P(|\psi|^{2})$ at small $|\psi|^{2}$:
\begin{equation}
\label{assimpt-small}
P(|\psi|^{2})=\frac{\Gamma^{4}(1/4)}{16\pi^{2}}\,\frac{\ell_{0}}{L}\,\frac{1}{|\psi|^{2}}=
\frac{\ell^{{\rm ext}}}{L}\,\frac{1}{|\psi|^{2}},\;\;\;\;({\rm
e}^{-L/\ell^{{\rm ext}}}\ll |\psi|^{2}\ell_{0}\ll 1).
\end{equation}
As was expected, the numerical coefficient in
Eq.(\ref{assimpt-small}) exactly corresponds to the replacement
$\ell_{0}\rightarrow \ell^{ext}$ in the leading term of expansion of 
Eq.(\ref{1d-dist} in accordance with Eq.(\ref{l-to-l0}).

However, finding sub-leading terms in $P(|\psi|^2 \rightarrow 0)$ 
is a separate non-trivial problem. We leave its complete study for 
future publications, outlining here only the origin of the difficulties. 
The point is that one can obtain the {\it formal} series of the type
Eq.(\ref{exp-an}) for $\Phi_{an}(u,\phi)$ with the coefficients
$g_{n}(\phi)$ and $f_{n}(\phi)$ represented by a two-fold integrals.
To this end we exploit again the integral representation
Eq.(\ref{W-GR-modified}) of the Whittaker functions. Plugging it
into the exact solution Eq.(\ref{canon-final}) we do the
$\lambda$-integration exactly using the well-known integral
\cite{GR}:
\begin{equation}\label{lambda-int-exact}
\int_{0}^{\infty}\frac{d\lambda}{\lambda^{2}}\,{\rm
exp}^{-\frac{a}{\lambda}-b\lambda}=2\,\left(
\frac{b}{a}\right)^{\frac{1}{2}}\,K_{1}(2\sqrt{ab}).
\end{equation}
The result is expressed through the two-fold integral:
\begin{eqnarray}
\label{Phi-x-ln}
&&\Phi_{an}(u,\phi)=\frac{2^{5/2}}{\Gamma^{4}(1/4)}\,\sqrt{u}\,{\rm
Re}\,\int_{1}^{\infty}\frac{dt_{1}}{(t_{1}^{2}-1)^{\frac{3}{4}}}
\int_{1}^{\infty}\frac{dt_{2}}{(t_{2}^{2}-1)^{\frac{3}{4}}} \\
\nonumber
&&K_{1}\left(\sqrt{u}\;\;\sqrt{\frac{\epsilon}{2}\ln\left(\frac{t_{1}+1}{t_{1}-1}\right)+
\frac{\bar{\epsilon}}{2}\ln\left(\frac{t_{2}+1}
{t_{2}-1}\right)}\;\;\sqrt{\bar{\epsilon} t_{1}\cos^{2}\phi+\epsilon
t_{2}\sin^{2}\phi} \right)\\ \nonumber &\times& \left
[\frac{\epsilon\ln\left(\frac{t_{1}+1}{t_{1}-1}\right)+\bar{\epsilon}\ln\left(\frac{t_{2}+1}
{t_{2}-1}\right)}{\bar{\epsilon} t_{1}\cos^{2}\phi+\epsilon
t_{2}\sin^{2}\phi}\right]^{1/2},
\end{eqnarray}
where $\epsilon=e^{i\pi/4}$, $\bar{\epsilon}=e^{-i\pi/4}$.

In Eq.(\ref{Phi-x-ln}) one can immediately recognize the combination
$\sqrt{u}\,K_{1}(\sqrt{u}...)$ which enters
Eq.(\ref{ord-solution}) and which generates the series
Eq.(\ref{exp-an}). The coefficients $g_{n}(\phi)=
g_{n}^{(1)}(\phi)+g_{n}^{(2)}(\phi)$ and
$f_{n}(\phi)=g_{n}^{(1)}(\phi)\,(f_{n}/g_{n})$ in the corresponding
series for $\Phi_{an}(u,\phi)$ are expressed in terms of the
coefficients $g_{n}$ and $f_{n}$ appearing in the expansion
Eq.(\ref{exp-ord}) of $2\sqrt{u}\,K_{1}(2\sqrt{u})$ and the two-fold
integrals:
\begin{eqnarray}
\label{ln-series-g} g_{n}^{(1)}(\phi)&=& c_{n}\,g_{n}\,{\rm
Re}\int_{1}^{\infty}\frac{dt_{1}}{(t_{1}^{2}-1)^{\frac{3}{4}}}
\int_{1}^{\infty}\frac{dt_{2}}{(t_{2}^{2}-1)^{\frac{3}{4}}}\\
\nonumber &\times&
[G(t_{1},t_{2})]^{n}\,[T_{\phi}(t_{1},t_{2})]^{n-1},\\
g_{n}^{(2)}(\phi)&=&c_{n}\,f_{n}\,{\rm
Re}\int_{1}^{\infty}\frac{dt_{1}}{(t_{1}^{2}-1)^{\frac{3}{4}}}
\int_{1}^{\infty}\frac{dt_{2}}{(t_{2}^{2}-1)^{\frac{3}{4}}}\\
\nonumber &\times& L_{\phi}(t_{1},t_{2})\,
[G(t_{1},t_{2})]^{n}\,[T_{\phi}(t_{1},t_{2})]^{n-1}\nonumber
\end{eqnarray}
of the three functions:
\begin{eqnarray}
\label{L-T}
&&G(t_{1},t_{2})=\epsilon\ln\left(\frac{t_{1}+1}{t_{1}-1}\right)+\bar{\epsilon}\ln\left(\frac{t_{2}+1}
{t_{2}-1}\right),\\  &&T_{\phi}(t_{1},t_{2})=\bar{\epsilon}
t_{1}\cos^{2}\phi+\epsilon t_{2}\sin^{2}\phi\\ &&
L_{\phi}(t_{1},t_{2})=\ln\left[G(t_{1},t_{2})\,T_{\phi}(t_{1},t_{2})/8\right],
\end{eqnarray}
where $c_{n}=\frac{2^{3(1-n)}}{\Gamma^{4}(1/4)}\,$.

One can check that $g_{0}\equiv g_{0}^{(1)}$ coincides with the
phase distribution function ${\cal P}_{an}(\phi)$ defined in
Eq.(\ref{P0-solution}). Furthermore, $f_{1}(\phi)$  appears to be
manifestly $\phi$-independent and coincides with $F_{1}$ found above
(see Eq.(\ref{F1})). The functions $g_{1}^{(1)}(\phi)$ (which is
also $\phi$-independent) and $g_{1}^{(2)}(\phi)$ are also well
defined.

However, starting from $n=2$   there is a problem in
Eq.(\ref{ln-series-g}). As the function $T_{\phi}(t_{1},t_{2})$
grows linearly with $t_{1,2}$, the integrals in
Eq.(\ref{ln-series-g}) are divergent for all $n\geq 2$. This signals
that the expansion Eq.(\ref{exp-an}) for $\Phi_{an}(u,\phi)$  breaks
down. The reason is that Eq.(\ref{exp-an}) does not guarantee the
correct, decaying at large $u$ behavior of the generating function
$\Phi(u,\phi)$. Thus an {\it additional series} in
$\Phi_{an}(u,\phi)$ may be required to cancel possible divergence at
$u\rightarrow\infty$ of the function obtained by the analytical
continuation of the series Eq.(\ref{exp-an}). As the result the
sub-leading term in the expansion of $\Phi_{an}(u,\phi)$ at small
$u$ is not proportional to $ u^{2}\,\ln u$ (as for the generating
functions away from the $E=0$ anomaly) but could be much larger. 
This anomaly deserves a separate investigation.

\section{Discussion and Conclusion}
\label{Conclusion} The goal of this paper was two-fold. The first
objective  was an asymptotic analysis of the exact solution
Eq.(\ref{canon-final}) for the anomalous (at the center-of-band
anomaly, $E=0$) generating function
$\Phi_{an}(u,\phi)$  at large values of $u\gg 1$. The corresponding
result is expressed by
Eqs.(\ref{F-cal-saddle-action}) and
(\ref{Phi-Wick-z-asymp-general})-(\ref{bb-phi}).

Knowing this asymptotic one can compute various quantities of
interest related with the local statistics of eigenfunction
amplitudes. The simplest one is the distribution function of the
eigenfunction amplitudes $P(|\psi|^{2})$ which behavior at large
$|\psi|^{2}$ gives an idea about the probability of anomalously
strongly localized states. To find this asymptotic form at the
center-of-band anomaly was our principal physical objective. We
managed to obtain the asymptotic expression for $P_{an}(|\psi|^{2})$ in a
compact form Eqs.(\ref{P-psi-anomal-answer}) and (\ref{C-phi-c}). The
result is a bit surprising, as it shows a re-entrant behavior
summarized in Fig.1, which points out on the two competing physical
phenomena behind it. Another indication of the same phenomena was
first found in our earlier works \cite{KY-PRB, KY-Ann} where we
noticed two different scales characterizing the moments of
$|\psi|^{2}$.

We also analyzed the asymptotic of $\Phi_{an}(u,\phi)$ at small
values of $u$. The leading term $\Phi_{an}(0,\phi)$ gives the
distribution function of phases ${\cal P}(\phi)$ \cite{Wegner,
Derrida} which is related with the distribution of scattering
phases. The next-to leading term $\propto u\ln u$ contains
information about the tail of the typical wave function. We have
computed this term and shown that it is compatible with
the exponential localization with the Lyapunov exponent found
in Refs. \cite{Wegner, Derrida}. We have also shown how the
sub-leading term $u\ln u$ results in the universal leading
behavior of $P(|\psi|)\propto |\psi|^{-2}$ at small $|\psi|^{2}$.

However, it appears that computing the further terms of expansion at
small $u$ (which contain information about the pre-exponential
behavior of the tail of localized eigenfunctions) is a non-trivial
problem which deserves further investigation.

\subsection*{Acknowledgments}
We appreciate numerous advices from Dima Aristov on the advanced use
of Wolfram Mathematica and a support from RFBR grant 12-02-00100 (V.Y.).
A part of work was done during visits of V.Y. to the Abdus Salam
International Center for Theoretical Physics, which support is
highly acknowledged.

\appendix\section{Generating function $\Phi_{an}(u,\phi)$ in the vicinity of the critical
angle $\phi_c$}

The saddle-point derivation of the large $u$ asymptotic of the anomalous generating function $\Phi_{an}(u,\phi)$ in the form (\ref{Phi-Wick-z-asymp-general}) with the angular
dependent pre-exponential function (\ref{A-phi}) is valid everywhere except for a close vicinity of the
critical angle $\phi_c$, where the saddle-point solutions $z_{\pm}(\phi)$ to
Eq.(\ref{F-cal-saddle}) are close to the branching point $1/\sin^2{\phi}$ 
so that the power expansion of the actions $\mathcal{F}_{\pm}(z,\phi)$ breaks down.

To find the width of the critical region of small $\delta\phi = \phi - \phi_c$,
we introduce the distances $\Delta_{\pm}$ between the
saddle-points $z_{\pm}(\phi)$ and the right branching point $z = 1/\sin^2{(\phi)}$ :
\begin{eqnarray}\label{Delta-z-definition}
\hspace{-1.5cm}
\Delta_{\pm} \equiv \frac{1}{\sin^2{\phi}} - z_{\pm}(\phi) \propto (\delta \phi)^2
\, .
\end{eqnarray}
The latter estimate follows from Eq.(\ref{z-crit}) in the vicinity of the
critical angle; the dependence $\Delta_{\pm} \sim (\delta \phi)^2$ is clearly
seen in Fig.\ref{z-1-2}.

The previous saddle-point approach to Eq.(\ref{Phi-Wick-z-asymp}) is justified
as long as the width of the saddle-point peaks
$$|z - z_{\pm}(\phi)| \sim
u^{-1/4}|\partial^2 \mathcal{F}_{\pm}(z,\phi)/\partial z^2|^{-1/2}
\sim u^{-1/4}(\Delta_{\pm})^{1/4}$$
is much smaller than the distance $\Delta z_{\pm}$ from the right 
branching point. It follows from here
that $|\delta \phi|$ should be greater than $u^{-1/6}$.

In the narrow region (of width $|\delta\phi | \leq
u^{-1/6}$) around the critical
angle $\phi_c$, the regular series expansion of the actions $\mathcal{F}_{\pm}(z,\phi)$
breaks down and the previous saddle-point approach is not applicable.
To treat this narrow critical region, we will develop a modified approach.
In fact, we will
restrict the analysis to even narrower vicinity of $\phi_c$: $|\delta \phi| \ll u^{-1/6}$,
which is sufficient for the calculation of the eigenstates distribution function $P_{an}(|\psi|^2)$ performed in the section \ref{Asymptotic of P}.

Note that the integrand in Eq.(\ref{Phi-Wick-z-asymp}) needs revision in the critical domain, too. This is because the saddle point estimate Eq.(\ref{I2-saddle}) for the integral $I_2(z,\phi)$ Eq.(\ref{I-definition}) does not work when the two saddle points $t_{\pm}$ Eq.(\ref{t-large-z}) approach each other (both go to zero at $z \rightarrow 1/\sin^2\phi $). The saddle-point estimate of the integral is valid only as long as the width of the saddle-point regions $|t-t_{\pm}| \sim u^{-1/4}|\delta z|^{-1/4}/\sin\phi$  is small as compared to the distance $|t_{+} - t_{-}| \sim |\delta z|^{1/2}\sin\phi$ between the two saddle points. Here we represented the integration variable $z$ in the form:
\begin{eqnarray}\label{z-in-criticality}
z = \frac{1}{\sin^2{\phi}} + \delta z
 \, .
\end{eqnarray}
From the above estimates one finds that the saddle-point approach is not applicable when
$|\delta z| \leq u^{-1/3}$ and therefore $|t_{\pm}| \leq u^{-1/6}$. For small $|\delta z | \ll 1$ the leading contribution to the integral $I_2[C] = I_2[C_{+}]$ (see Eq.(\ref{I-C-plus}) and the discussion around it) comes from a narrow vicinity of the origin $t=0$. Expanding
the function $f_2(t,z,\phi)$ Eq.(\ref{f2}) in small $t$ (and keeping only linear terms
in $\delta z$), we arrive at the following expression for $I_{2}[C]$:
\begin{eqnarray}\label{I2C-in-criticality}
I_2[C] = \mathrm{e}^{-\frac{3\pi}{4}i} \mathrm{e}^{-\frac{\pi\sqrt{u}\sin{\phi}}{4}} \int^{\infty}_{-\infty} dt \, \mathrm{e}^{i\frac{\sqrt{u}\sin{\phi}}{2}\left[\frac{t^3}{3} - t \delta z \sin^2{\phi} \right]}
 \, .
\end{eqnarray}
This expression and the relation Eq.(\ref{I-C}) determine the quantity of our interest, $\Re \left[\mathrm{e}^{i\pi/4}I_2(z, \phi) \right] $; the integral representation (\ref{Phi-Wick-z}) for the anomalous generating function $\Phi_{an}(u, \phi)$  takes the form:
\begin{eqnarray}\label{Phi-Wick-z-critical}
&&\Phi_{an}(u,\phi) = \frac{2^{1/3}(2\pi)^{3/2}\sin^{2/3}{(\phi)} \,\, u^{1/12}}{\Gamma^4(1/4)}
\mathrm{e}^{-\sqrt{u}\, [\cot{(\phi)} - \sin{(\phi)} \,\, Y(\phi)/2]} \nonumber \\
&& \int^{\infty}_{-\infty} d(\delta z) \,
\mathrm{e}^{-\sqrt{u} \,\, \delta z \sin^3{(\phi)} \,\, Y(\phi)/4}\,
\mathrm{Ai}\left(-2^{-2/3}\sin^{8/3}{(\phi)} \, u^{1/3} \delta z\right)
\, ,
\end{eqnarray}
with $\mathrm{Ai}(x)$ being the Airy function.
According to Eq.(\ref{crit-angle}) the function $Y(\phi) \sim \delta \phi$
in the critical region, so the linear in $\delta z$ term in the exponent
of the integrand can be safely omitted.
Indeed, due to the convergence of the integral of the Airy function,
an estimate for typical $\delta z$ in the integral Eq.(\ref{Phi-Wick-z-critical})
is $\delta z \sim u^{-1/3}$, hence the linear in $\delta z $ term in the exponent is estimated as $u^{1/2}\delta\phi \, \delta z \sim u^{1/6}\delta\phi$ which is negligible
in the considered critical region $|\delta \phi| \ll u^{-1/6}$.
Neglecting the linear in $\delta z$ term in the exponent and calculating the remaining
integral of Airy function, we obtain $\Phi_{an}(u, \phi)$ in the vicinity of the critical
angle $\phi_c$ in the form Eq.(\ref{Phi-Wick-z-asymp-general})
with the pre-exponential function $A(\phi) \rightarrow A_c(\phi)$:
\begin{eqnarray}\label{A-phi-crit}
\hspace{-1.5cm}
A_c(\phi) = \frac{2(2\pi)^{3/2}}{\Gamma^4(1/4)\sin^2{\phi_c}}
\, .
\end{eqnarray}
This expression matches perfectly the out-of-critical expression
Eq.(\ref{A-phi}) when the latter is formally extended to the critical region
$\phi \approx \phi_c$, where $z_{\pm}(\phi) \approx 1/\sin^2\phi$. This comparison
completes our derivation of the asymptotic of the generating function
$\Phi_{an}(u, \phi)$ both in and out of the ``critical region" and shows that
$\Phi_{an}(u, \phi)$ is a smooth function of $\phi$ even in the vicinity of the ``critical
angle" $\phi_c$.


\begin{thebibliography}{99}
\bibitem{box-distribution} For the box-like probability distribution of the site disorder
$\rho(\varepsilon)=\theta(W/2 -|\varepsilon|)/W$, one has $\sigma^2 = W^2/12$ and $\ell_{0} = 96/W^2$.
\bibitem{rem} The physical meaning of the restriction in the brackets  will
be explained in Sec.4.
\end{thebibliography}
\end{document}